\documentclass[aps,prb,amsmath,amssymb,amsfonts,twocolumn,nofootinbib]{revtex4}
\usepackage{graphicx}% Include figure files
\usepackage{dcolumn}% Align table columns on decimal point
\usepackage{bm}% bold math
\usepackage{amsmath}
\usepackage{amssymb}
\usepackage{color}

\newcommand{\be}{\begin{equation}}
\newcommand{\ee}{\end{equation}}
\newcommand{\bea}{\begin{eqnarray}}
\newcommand{\eea}{\end{eqnarray}}
\newcommand{\pp}{\partial}

\newcommand{\bma}{\begin{matrix}}
\newcommand{\ema}{\end{matrix}}

\begin{document}

%\title{Measuring eigenfrequency distribution of graphene membrane randomly attached to the substrate via phonon assisted Tien-Gordon effect}

\title{Eigenfrequencies of the randomly pinned drum and conductivity of graphene}

\author{M. V. Medvedyeva}
\affiliation{Instituut Lorentz, Leiden University, Niels Bohrweg 2,2300 RA Leiden, Netherlands}
\affiliation{Department of Physics, Georg-August-Universit\"{a}t G\"{o}ttingen,
Friedrich-Hund-Platz 1, 37077 G\"{o}ttingen, Germany (present)}
\author{Ya. M. Blanter}
\affiliation{Kavli Institute of NanoScience, Delft University of Technology,
Lorentzweg 1, 2628 CJ Delft, The Netherlands}

\begin{abstract}
Graphene is convenient material for nanomechanichal applications since high-frequency oscillations are easily accessible. In this Article, we consider graphene on a rough substrate attached to imperfections at random locations.
We explore the statistics of low-lying phonon modes, which exert most influence on the conductivity of graphene. We find that {\it the nearest neighbor spacings of low lying eigenfrequencies have the Wigner-Dyson probability distribution after averaging over the random configurations of disorder}.
Due to interaction of electrons with the oscillations of the membrane, an electron can be transfered to higher or lower energies, which is a manifestation of the phonon-assisted Tien-Gordon effect.
The Tien-Gordon effect suppresses the conductivity of graphene.
In the regime of low Fermi energies and small sizes of the sample an increase of conductivity is observed which we refer to Klein tunneling and electron-hole pair creation.
Eventually, when the increase of the transmission becomes too prominent,
the pair creation changes the ground state of the system, signalizing the limit of applicability of the single-particle Dirac equation used in this paper.
\end{abstract}


\maketitle

\section{Introduction}

Graphene is a monolayer of graphite with excellent elastic properties and low mass, which make it a prospective nanomechanical resonator~\cite{bunch} attractive for a wide range of applications from mass sensing in single-molecule range~\cite{mass} to quantum manipulation of the elementary mechanical vibrations~\cite{quantum}. They are combined with perfect electrical properties. The mobility of the carriers, both in suspended samples~\cite{suspended} and the samples on the substrate~\cite{BN}, is one of the highest among all modern semiconductors. The combination of mechanical and electrical properties opens wide opportunities for manipulation and read-out of mechanical oscillations using electrical signals.

In most experiments, graphene is deposited on substrate. The substrate is often not ideal and can be considered as a random potential landscape. The graphene membrane interacts with the substrate, which may result to its attachment to the substrate at some areas or points, and free suspension between these areas. This arrangement was confirmed experimentally~\cite{Morgenstern}. This means that the mechanical oscillation spectrum of such membrane is complex, with localized and extended phonons determined by the substrate profile. Moreover, since the profile is random, the frequencies of these excitations can be considered as random quantities. It is known that the frequencies do not uniquely characterize the shape of the oscillating membrane. Indeed, this question was studied in 1966 by Marc Kac in his famous paper "Can one hear the shape of the drum"\cite{Kac},
%\cite{Weyl},
with the conclusion that is possible to construct different in shape, but isospectral membranes. Therefore the statistical approach remains the only meaningful way to characterize the frequency spectrum of this system.

In this Article, we characterize statistical properties of these frequencies and show how they can be assessed via electric transport measurements.

The elastic energy of the oscillating membrane has two contributions: stretching and bending energies. We conclude that the bending energy is much smaller than the stretching one. Hence the dynamics of graphene membrane is described by the Helmholtz's equation with the fixed boundary conditions at the regions where the membrane is attached to the substrate. The same equations describes the oscillation of the classical membrane. In our work we show how to "hear" the disorder of attachments of the membrane and see the distribution of the low frequencies averaged over the disorder. The frequencies can be directly measured in the experiment, but for graphene it is more convenient to extract information on the oscillation of the membrane from the conductivity measurements.

The high-energy spectrum of the eigenvalues of the Laplace operator is well studied and is known to obey the Wigner-Dyson distribution~\cite{Haake}. This statement was proven by a variety of methods, analytical (random matrix theory and non-linear sigma-model) as well as numerical (various models of quantum billiards). Therefore in this Article we study the probability distribution of the low energy spectrum of the membrane on disordered substrate. This question was not previously addressed since most of the studies of eigenvalues concern electron systems, where the relevant energy levels lie around the Fermi surface and have very high energy. In contrast, the phonon frequencies have a clear cut-off at low energies. This cut-off frequency determines the scale for low-frequency modes. The second reason why low energy levels are beyond the scope of the usual methods is that no statistical analysis can be performed for only few levels. Thus, for a clean system the problem of the statistics of low-lying eigenvalues can not be formulated. One needs a disordered system to address this issue: Averaging over the disorder acts as averaging over the statistical ensemble. The problem can be addressed only numerically as the analytical methods mentioned above are not applicable for low energy spectrum. According to our numerical modeling {\it the distribution of low frequency eigenvalues of the membrane randomly attached to the substrate also reveals Wigner-Dyson statistics}.

Electrons in graphene are coupled to the deformation of the lattice. Due to the deformation of the lattice the overlap between the electron orbitals is changed, so that the band structure which describes the effective motion of the electrons is altered. Therefore the electron conductivity can be influenced by the oscillations of the membrane. In monolayer graphene, the deformations can be strong, which leads to strong electron-phonon coupling.
The coupling of the out of plane phonons to the electrons results in electrostatic potential and pseudomagnetic field which depend on the deformation, acting on electrons~\cite{Guinea}. Since electrons move in the oscillating potential created by the phonons, they can scatter with absoprtion or emission of energy from the membrane. If the current-voltage characteristic of the system has some structure at the voltage $V$,  then this structure propagates also to the voltages   $V+n\hbar \omega_i$, with $\omega_i$ being the set of the membrane frequencies and $n$ an integer number. This phenomenon is similar to the Tien-Gordon effect which arises if the one lead of the sample is irradiated by the photons~\cite{Tie63}. In the situation we consider, however, the oscillating potential is created by the membrane itself, and the area where the electrons feel this potential is separated from the leads by potential barriers. Then the conductance exhibits Fabry-Perot resonances at the energies $E_l$. Due to the oscillations with frequencies $\omega_i$ each peak is enhanced at the energies $E_l+ n\hbar \omega_i$. If the conduction modes of graphene at different frequencies do not interact, the presence of the addition to the Fabry-Perot peaks provides us with an opportunity to measure the distribution of eigenfrequencies of the oscillating membrane.
This way of detecting the oscillations of the membrane will work for narrow graphene strips. For wide strips, the mixture of Fabry-Perot peaks from different harmonics and their satellites will not give a clear picture.

A number of papers~\cite{Silvia1,Silvia,VonOppen,Che08,Akt08,Kachorovski} studied phonons in graphene. They were focused on phonon spectrum either not influenced by disorder or influenced by weak disorder. We consider the special case, when the disorder changes macroscopic motion of the membrane, namely, the graphene sheet is attached to the substrate at random points.

This Article is structured as follows. In Section~\ref{secmodel} we present the detailed results on the statistics of low-lying eigenfrequencies of the graphene membrane. The numerical method used to obtain the results is found in Appendix~\ref{BoundaryIntegral}. In Section~\ref{phonTG} the conductivity modification due to the Tien-Gordon effect is considered first for one frequency, then for two frequencies, and finally the influence of the distribution of the phonon modes on the conductivity is addressed. In Section~\ref{discuss} we discuss the results.
In Appendix~\ref{AppendixTG} we show the method used to solve the scattering problem for determining the conductivity of the oscillating membrane.

\section{Phonons in disordered graphene}
\label{secmodel}
Graphene on corrugated substrate can be modeled as a membrane which is randomly attached to some regions of the substrate. We describe this membrane using the boundary integral method and confirm the usual wisdom that the distribution of the spacings between the nearest neighbour phonon energy levels for high levels is the Wigner-Dyson distribution. Moreover, we investigate the statistics for the low energies and find that, after averaging over disorder, it is also described by the Wigner-Dyson type function.

\subsection{Model}

\begin{figure}[ht!]
\begin{center}
\includegraphics[width=0.95\linewidth]{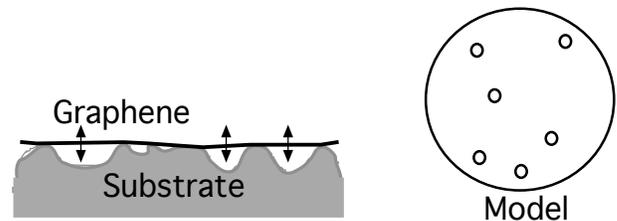}
\caption{\label{model}
Left: Schematic representation of graphene attached at the arbitrary regions to the substrate.
Graphene can oscillate between the regions of attachment.
Right: The model used to calculate the frequencies of oscillations of graphene membrane: circular membrane with circular regions of attachments placed randomly.}
\end{center}
\end{figure}

Scanning tunneling miscoscopy (STM) experiments show\cite{Morgenstern} that graphene on a rough substrate is attached to the substrate at some regions, at the hills of the substrate, and it is freely suspended over the valleys. To take this into account, we model a graphene sheet on a substrate as a membrane which is fixed at some regions. Due to external forcing or due to interaction with environment the membrane can vibrate. In this Section, we are interested in the statistics of eigenfrequencies of such membrane.

We assume that the membrane is attached to the substrate strongly enough, so that it can not be detached due to oscillations. The depinning of the membrane on the rough substrate was considered in Ref.~\onlinecite{Silvia}. The authors concluded that for the large scale of the imperfections of the substrate the membrane indeed remains attached to the substrate, namely for $h/\mathbf{h}<\mathbf{h}\sqrt{\delta_c/\kappa}(l/\mathbf {h})^2$, where $h$ is the height of the fluctuations, $l$ is the characteristic length of the fluctuations, at which the attachment occurs, $\delta_c$ is the coupling strength of the graphene to the substrate, estimated for graphene in $\text{SiO}_2$ as $\sim2$meV~\cite{gammas}, $\kappa$ is the bending rigidity of the two-dimensional layer, $\kappa\sim1$eV, and the length scale $\mathbf{h}$ is the effective thickness of graphene layer $\mathbf{h}=\sqrt{\kappa/E_{2D}}$, $\mathbf{h}\approx1\AA$, $E_{2D}$ is the two-dimensional Young modulus, $E_{2d} =340 N/m$. For example, if the height fluctuations are of the order of $5\AA$, the attachment region of $1$nm is sufficient to keep the membrane attached to the substrate. We assume below that the condition is met, and depinning does not occur.

The Lagrangian of the oscillating membrane is
\be \mathcal{L} = \frac{\rho_0 \dot{\phi}^2}{2}-\frac{1}{2}\kappa(\nabla^2 \phi)^2-\frac{1}{2}\delta (\nabla \phi)^2, \label{Lagr}\ee
with $\phi(t)$ being out-of-plane deviation. The first term in Eq. (\ref{Lagr}) stands for the kinetic energy, $\rho_0$ is the density of the two-dimensional graphene layer, $\rho_0=7.6 \times 10^{-7} kg/m^2$, the second term is the bending energy.
The third term is the energy of the deformed membrane with the tension $\delta$, which can be estimated via the Hooke's law $ \delta= E_{2d}\Delta L/L$ with the relative elongation $\Delta L / L = \delta_0 / E_{2d} + \xi / L $. Here $\delta_0$ is the initial pretension and $\xi$ is the deformation of the layer (the amplitude of the oscillations).
Hence we get the ratio of the bending and strectching energies $E_{bend}/E_{str} \sim (\mathbf{h}/L)^2 (\Delta L / L)^{-1}$. For a large membrane, this ratio is small, since, $\mathbf{h}/L \ll 1$.
In the following, in order to obtain the eigenfrequencies of the membrane, we only take into account the term determined by stretching.

Assuming that the oscillations are harmonic, $\phi \propto \exp(i \omega t)$, we obtain the equation for the spectrum of the membrane with the border $\Omega$  (at which the out-of-plane displacement vanishes),
\bea \Delta \phi(q) + \lambda_n^2 \phi(q) =0, \text{ for q in }\Omega/\pp\Omega \ , \label{laplas} \\
 \phi(q) =0 \text{ for $q$ in } \pp \Omega \ ,\label{bounlaplas}\eea
with $\lambda_n^2= \rho_0 \omega_n^2/\delta$. The boundary
$\Omega$ consists of the external border of the  graphene sheet $\Omega_0$ and of the areas $\{ \Omega_1, \Omega_2,\ldots,\Omega_N \}$ where graphene is attached to the substrate.
We note that formally $\lambda_n^2$ correspond to the energy in the equivalent Schr\"odinger equation widely discussed in the context of quantum chaos. Therefore when we consider the statistics of the eigenfrequencies of the oscillations we always refer to $\lambda_n^2$, not to $\lambda_n$.

The eigenvalues of the system~(\ref{laplas}),~(\ref{bounlaplas}) are found numerically by using the boundary integral method\cite{Baecker}, see Appendix~\ref{BoundaryIntegral} for more detail. This method is based on the re-writing the eigenvalue problem for the Laplace equation~(\ref{laplas}) and its boundary condition~(\ref{bounlaplas}) in the form of the integral equation for the normal derivative of $\phi$ on the boundary. It leads to solving the eigenvalue problem for the system of linear equations in a numerical implementation of the method. We generalized the method for the case of several boundaries (see Appendix~\ref{BoundaryIntegral}). We tested the program on exactly solvable problems such as an eigenfrequencies of a circle and of a circle with a hole in the middle.

To model graphene on substrate we take a circle with circular regions of the attachments in the middle (to be referred below as impurities or holes). The model is general enough as its classical analogue is non-integrable already if two circles are randomly place inside the large circle.
The impurities are placed at random positions. The eigenfrequencies are computed for each configuration. The averaging over disorder configurations is performed to determine the statistics.

\subsection{Statistics of eigenfrequencies at high energies}

\begin{figure}[ht!]
\begin{center}
A\includegraphics[width=0.95\linewidth]{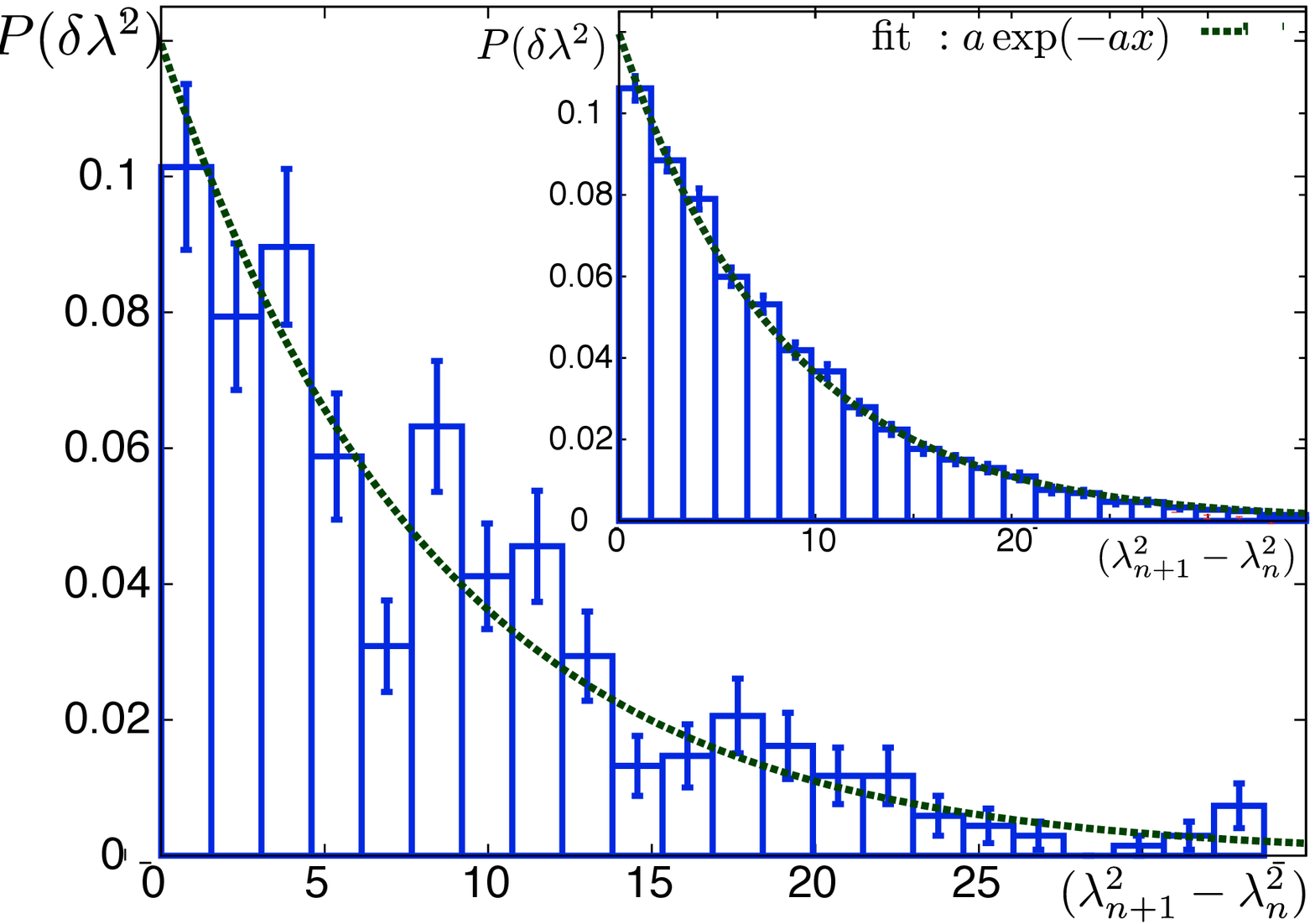}
B\includegraphics[width=0.95\linewidth]{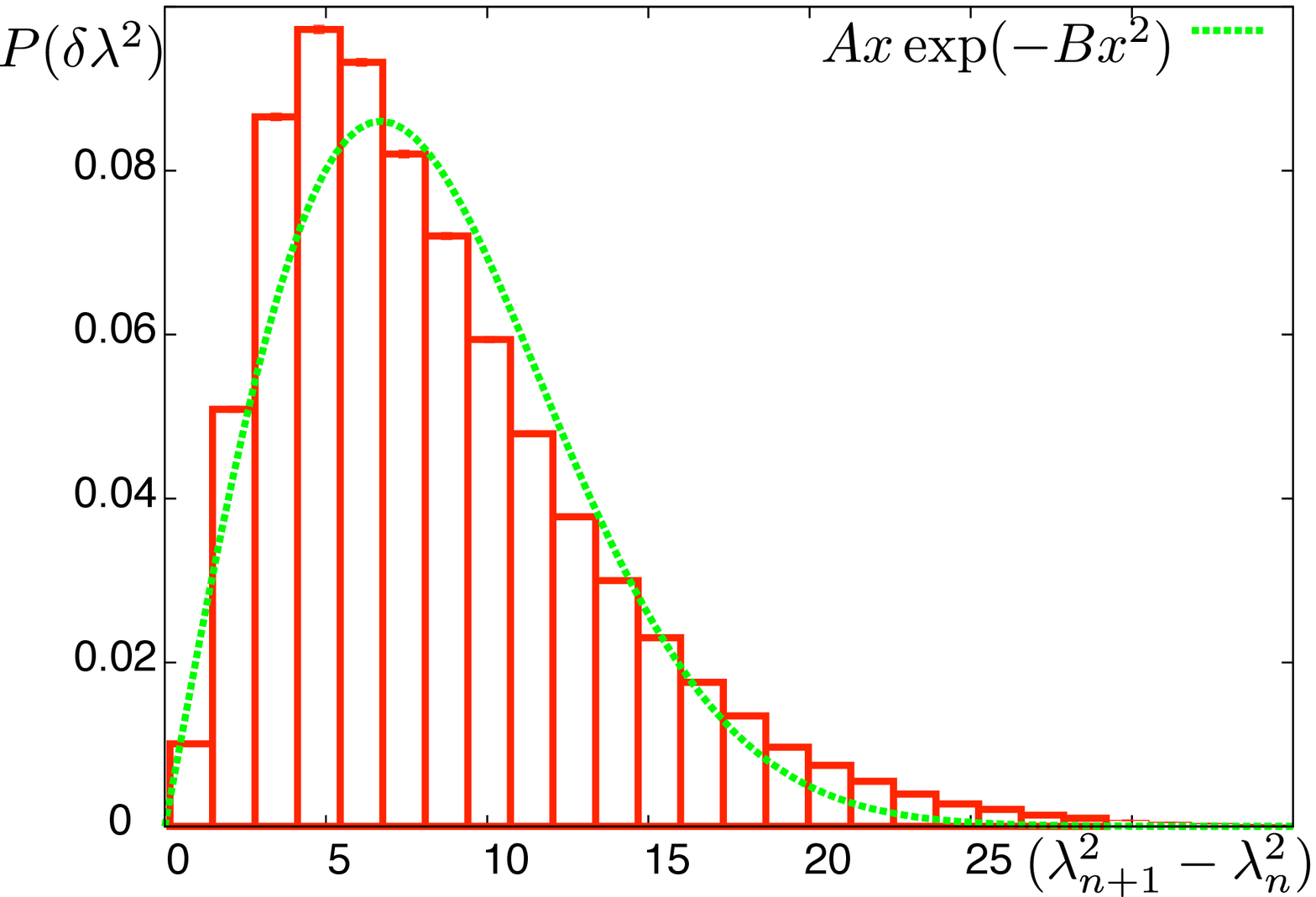}
\caption{\label{long}
The distribution of the nearest neighbour levels spacings is shown for the integrable case of the circular membrane and for the case of the membrane with holes. The distributions are taken for the same range of the energy levels. For the integrable circular membrane the distribution (A) reveals the Poissonian distribution function~(\ref{Pois}), $N=444$. We note that the distribution is quite noisy. There is no ensemble average for the integrable system, hence for the fixed range of energies there is a fixed number of levels. In the inset we show the distribution for larger range of energy, $N=7180$, which is well fit by the Poissobian distribution. For the chaotic membrane (B) the distribution of the high energy levels reveals Wigner-Dyson probability distribution~(\ref{Wigner}). The distribution is computed for the two randomly placed holes. The number of the levels is $N=401442$.}
\end{center}
\end{figure}

Without any regions of the attachment in the middle the system (a circle) is integrable, and its eigenfrequencies obey Poissonian statistics,
\be P(x)\sim\exp(-x), \ \ x = \lambda_{n+1}^2 - \lambda_n^2 \ . \label{Pois}\ee
see Fig.~\ref{long}B.
A circle with two impurities in the middle becomes classically chaotic, namely a small initial difference between trajectories grows infinitely in time. It is well-known that the statistics of high-energy levels of classically chaotic systems (such as non-integrable billiards) is well fit by the same distribution function~\cite{Mir00,Haake}. The same statistics appears in the disordered solid state problem treated by non-linear sigma-model, see Ref.~\onlinecite{Mir00}.
Indeed, we find that for the high levels the statistics of the nearest level spacings is Wigner-Dyson,
\be
P(s)= \frac{\pi}{2}s\exp(-\tfrac{\pi}{4}s^2) \ ,
\label{Wigner}
\ee
with $s$ being the normalized spacing between two levels,
see Fig.~\ref{long}A.

To get higher eigenvalues, we need to discretize all boundaries very accurately, and it becomes computationally consuming to implement more holes. This is why  we have chosen the minimal number of impurities (two) for which the system is chaotic.

We note that if one models the attachment regions as points, the system is classically integrable, hence the statistics of the levels for a correctly chosen window of the energies is Poissonian~\cite{Tudorovsky}, even though for a smaller window it is Wigner-Dyson. In this manuscript, we remain in the Wigner-Dyson regime.

\subsection{Statistics of low eigenfrequencies of the membrane}

\begin{figure}[ht!]
\begin{center}
\includegraphics[width=0.95 \linewidth]{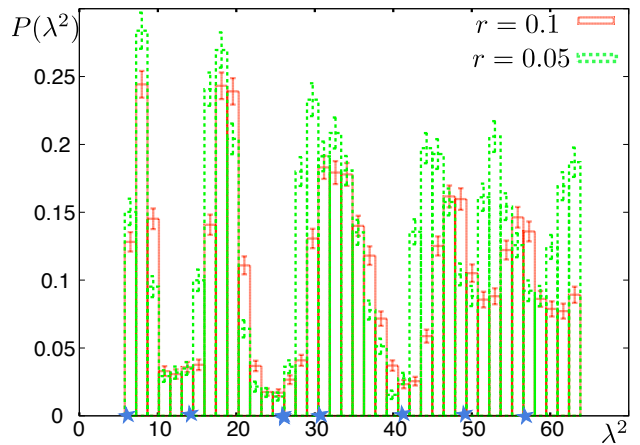}
\label{density_osc}
\caption{The density of eigenfrequencies for the first few oscillation modes of the membrane. The density is smeared by disorder in the position of the holes, but the oscillating behavior reflecting the positions of the eigenvalues in the absence of the disorder is present. We notice that in the region of the fourth-fifth initial eigenvalue of the membrane the density of states is already substantially more smooth that for the first few levels.
The data is shown for two different sizes of the holes, $r=0.1$ and $r=0.05$. The averaging is done for the different number of the holes $N=5,6,7$.
}
\end{center}
\end{figure}

\begin{figure}[ht!]
\begin{center}
A\includegraphics[width=0.95\linewidth]{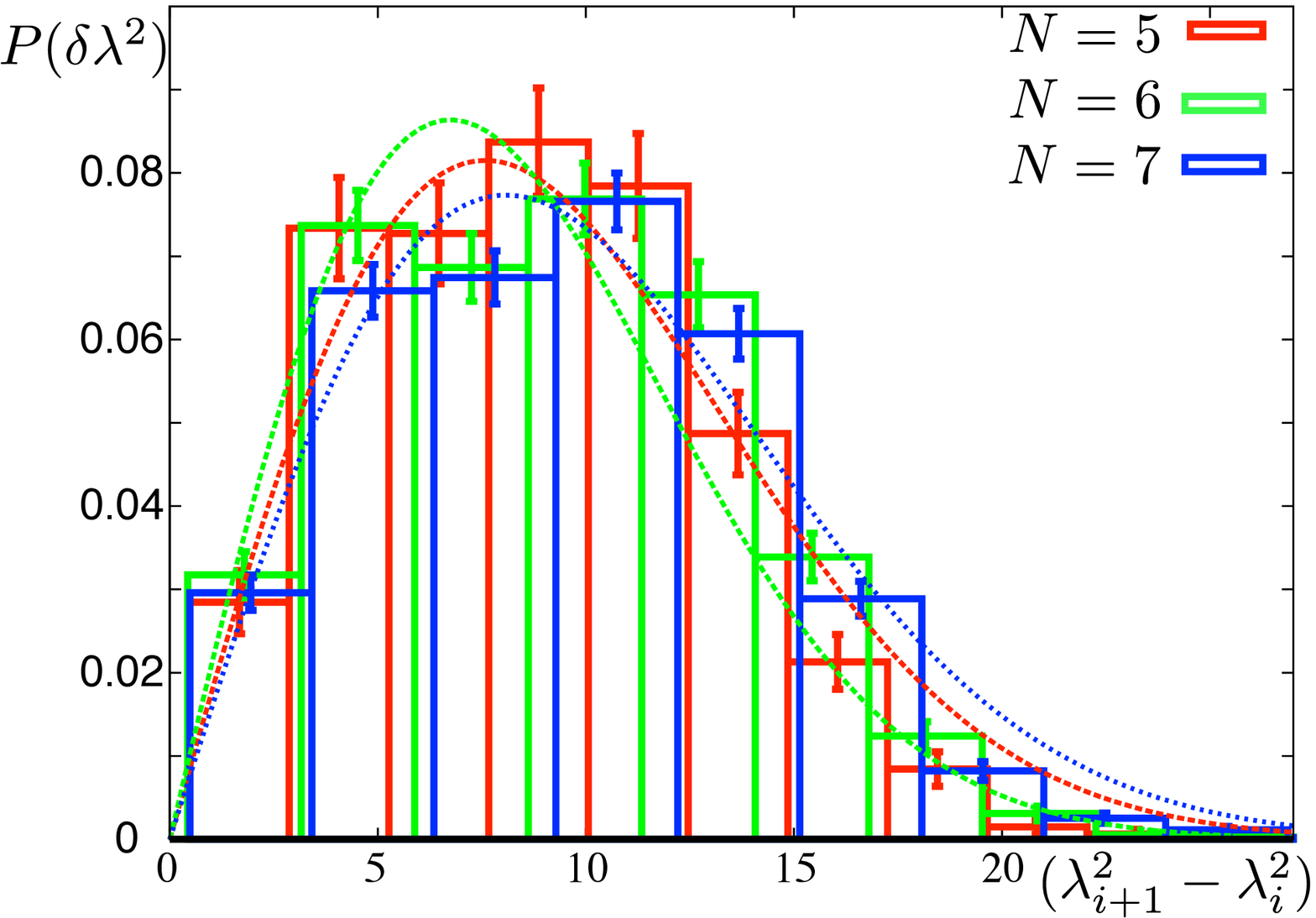}
B\includegraphics[width=0.95\linewidth]{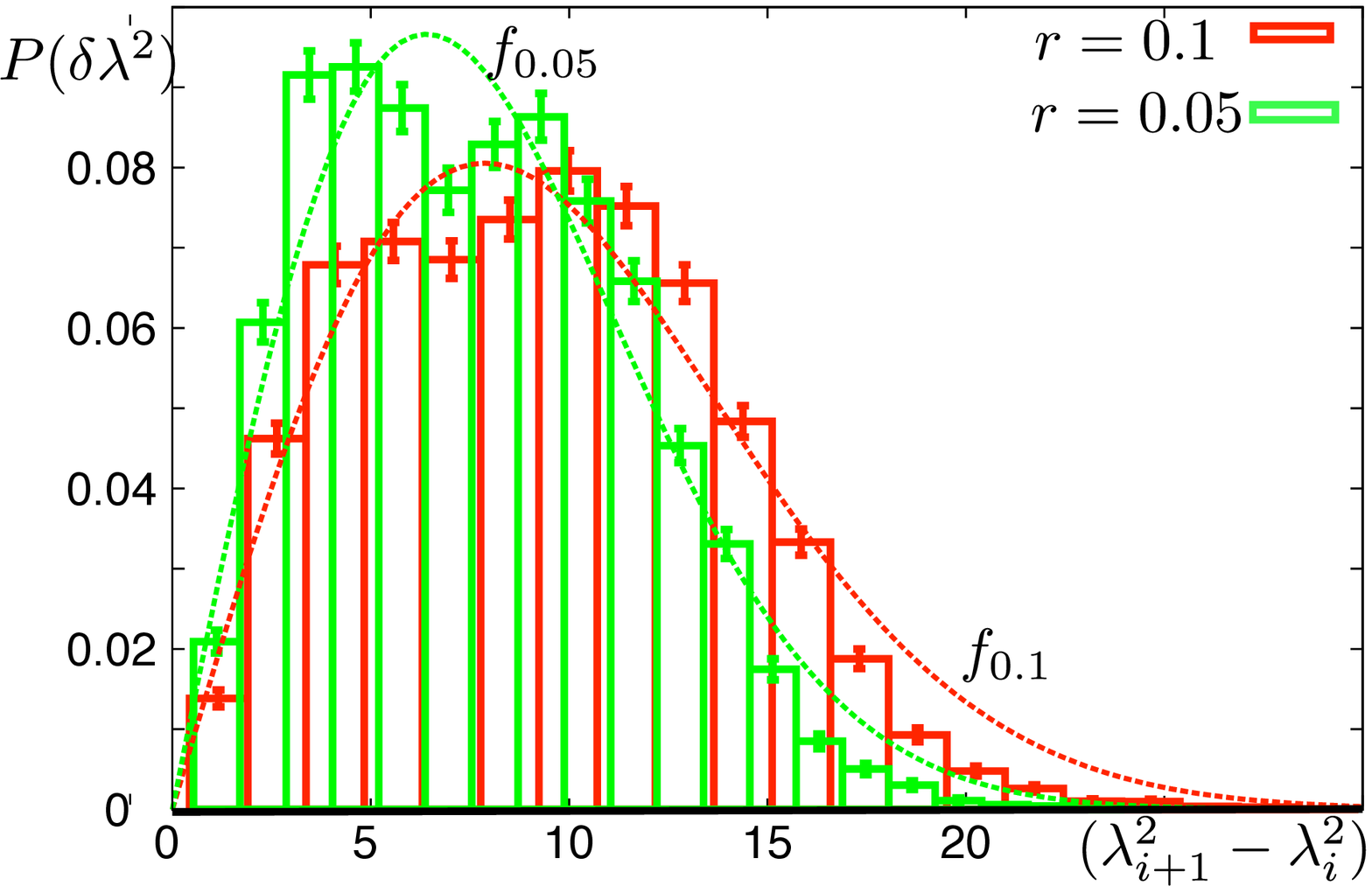}
\caption{\label{NNL}
Statistics of the low-lying eigenfrequencies.
A: The distribution of the NNL (next nearest levels) for different numbers of impurities, $N=5,6,7$. The distributions has similar properties as we see from the histograms and from the fit.
B: The distribution of the NNL for the first few eigenfrequencies for different size of the impuritites, $r=0.1$ and $r=0.05$ (averaging over different number of impurities, $N=5,6,7$). We notice that it reproduces well the Wigner-Dyson distribution function~(\ref{Wigner}). The normalization is done for the averaged to the constant density of states.
}
\end{center}
\end{figure}

\begin{figure}[ht!]
\begin{center}
A\includegraphics[width=0.95\linewidth]{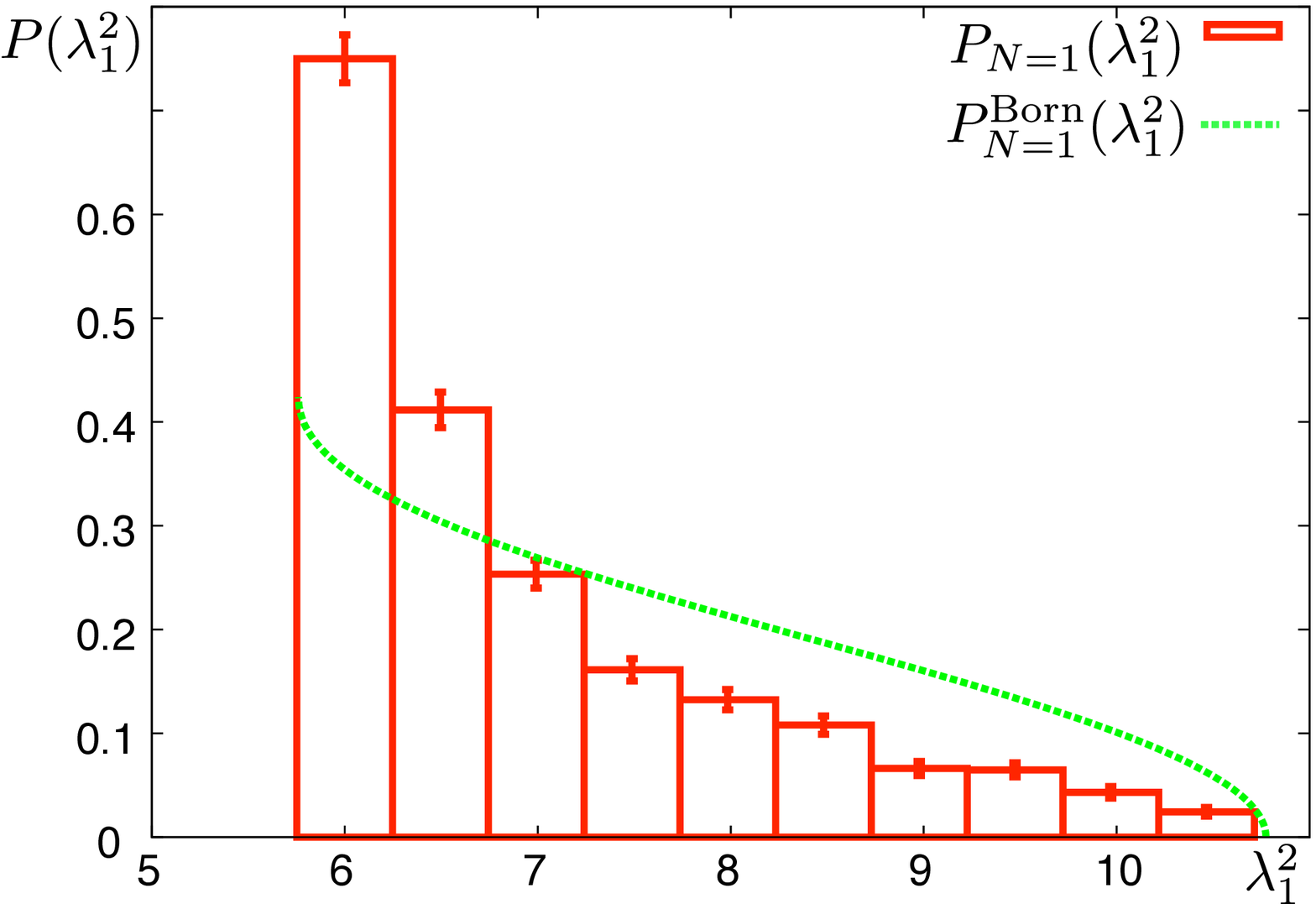}
B\includegraphics[width=0.95\linewidth]{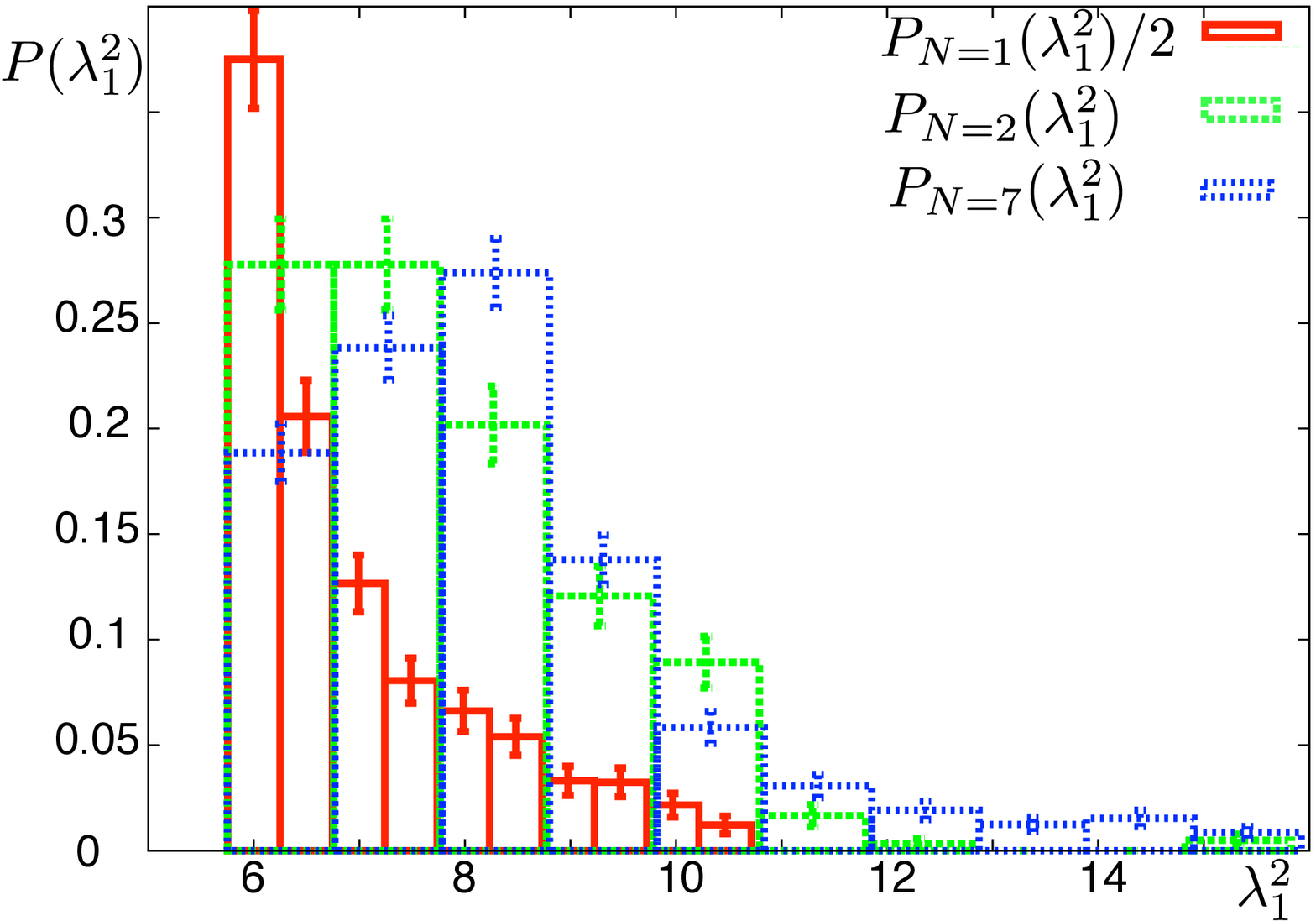}
C\includegraphics[width=0.95\linewidth]{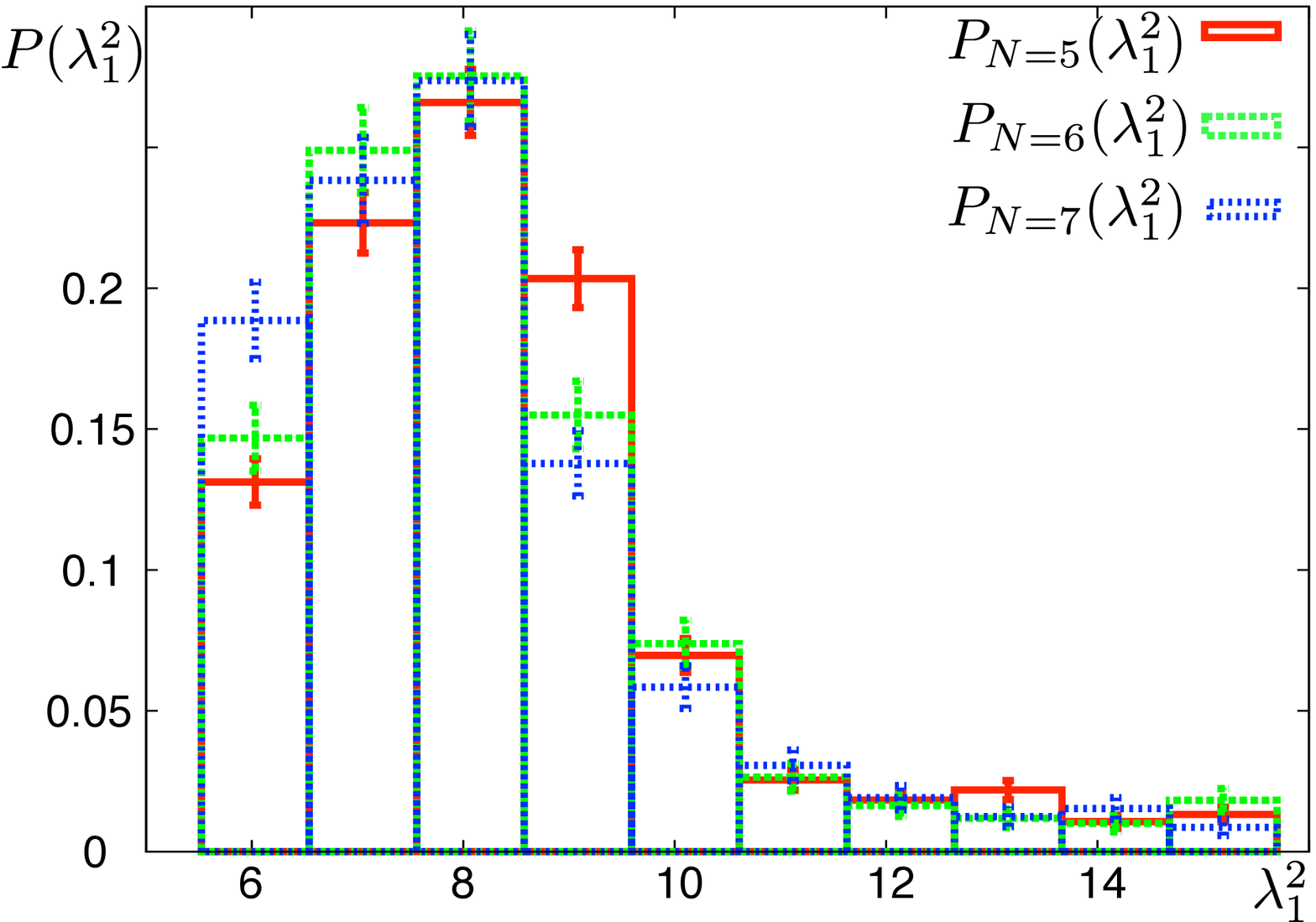}
\caption{\label{first_level}
The energy distribution of the first eigenfrequency, $P_N(\lambda^2_1)$, for different numbers of the attachement regions, $N$.
A: $P_1(\lambda_1^2)$ in comparison with the distribution in the Born approximation~(\ref{Born}),~(\ref{BornDistr}).
B: Evolution of the energy distribution increasing number of holes, $P_{1,2,7}(\lambda_1^2)$.
C: The energy distribution becomes stable for the large number of holes, $P_{5,6,7}(\lambda_1^2)$.
}
\end{center}
\end{figure}

The statistics of high-energy levels for our system is not different from any other system described by the Laplace equation. What is particular in a membrane over substrate is that we now have an access to statistics of low-lying levels as well, whereas for electron levels, which typically are the subject of investigation in the level statistics problems, low-lying levels are not accessible and are outside the scope of the problem.
For high levels, the random matrix theory appears as collective dynamics of the levels, and the statistics is effectively described by the Wigner-Dyson distribution. For low levels we do not have such averaging over the level positions.
To the best of our knowledge, statistics of low-lying levels has not been studied so far since the absence of large number of levels means that there is no ensemble over which one might perform averaging to reproduce the results of random matrix theory (which is where the Wigner-Dyson distribution comes from \cite{randommatrix}). It is not obvious what distribution the lowest lying levels obey.

We study this statistics in more detail than the one of high energy levels for different numbers of impurities. We treat the impurities as quenched disorder, hence we always have average over disorder. The density of states averaged over different disorder configurations oscillates revealing the position of zeros of the clean membrane and becomes  constant for large energies, see Fig.~\ref{density_osc}. We averaged over configurations with different numbers of the impurities, $N=5,6,7$. The oscillation in the density of levels are more pronounced for smaller size of the impurities (the number of impurities is the same). It is in agreement with the simple-minded argument that small impurities have less influence on the spectrum. We argue that that the oscillations in the density of low energy states do not disappear for the case of non-integrable boundary as the oscillations simply reveal the geometric shape of the membrane. Therefore for a non-integrable outer boundary of arbitrary shape the oscillations would be shifted according to the first few energy levels of the membrane without disorder.

We obtain the Wigner-Dyson distribution for the level spacings of the first four-five low lying energy levels slightly depending on the number of impurities in the system, Fig.~\ref{NNL}A. We average all obtained numbers for different number of impurities and get the distribution which is close to the Wigner-Dyson one for different sizes of the impurities, Fig.~\ref{NNL}A,B. Let us note that we fit the numerical data with the dependence which is the same as in Wigner Dyson probability distribution. In the Wigner-Dyson probability distribution for high energy levels, the parameter $s$ of the Wigner-Dyson probability function is normalized by the average level spacings $\delta \lambda^2$, namely,
$s=(\lambda_{n+1}^2-\lambda_n^2)/\delta \lambda^2$.
In the case of low levels, the mean level spacing for the region of fit oscillates, see~Fig.~{\ref{density_osc}}, the fit in Fig.~\ref{NNL} is close to the Wigner-Dyson fit with the parameter $s$ normalized by the mean value of density in the region of fitting.

Furthermore, we explore the statistics of the position of the first energy level in the chaotic disordered cavity,~Fig.~\ref{first_level}.

First we compare the distribution of the energy levels for one impurity obtained by solving the corresponding equation numerically and computed from the Born approximation. We show the difference between conventional way of computing the shift of eigenenergies due to the disorder for the electron system with an impurity of finite potential $V$ and the phonon system with fixed to zero boundary condition on the circumference of the impurity.

In the Born approximation, the shift of the energy of the first level for the impurity $\Omega_i$ cdplaced at the random position at the distance $\rho$ from the center is
\be E_1(\rho)-E_1^{(0)} = V \int_{\Omega_i} J_0^2(E_1^{(0)} \mathcal{r}) d\Omega_i \approx V \pi r^2J_0^2(E_1^{(0)}\rho) \ , \label{Born}\ee
where $V$ is the potential characterizing the degree of attachment of the membrane to the substrate, and $J_0$ is the Bessel function of the zeroth order, which represents the solution of the Laplace equation in a circle with the zero boundary conditions.
The probability to find the impurity at the distance $\rho$ from the center is
\be P(\rho)= 4 \pi r \rho. \label{BornDistr}\ee
We can get the dependence $\rho(E_1)$ from Eq.~(\ref{Born}) and consequently the probability distribution $P_1(E_1)$ in the Born approximation.
Fig.~\ref{first_level}A demonstrates that the Born approximation does not correctly describe the problem with the zero boundary conditions. The discrepancy can be attributed to the fact that we need an infinite potential to reproduce zero boundary conditions at the finite circumference of the attachement region, while the Born approximation is only applicable for $V r^2 \ll 1$. For this reason, we proceed with numerical solution for a more complex system, with two and more impurities.

As we would expect from simple-minded considerations analogous to those in the quantum mechanics, the eigenfrequency can move only up in energy with increasing the amount of disorder (amount of impurities). Thus, the minimum eigenfrequency corresponds to the one of the ideal circle. On all our plots we display the distribution of the first energy level taking into account this finite energy offset. The distribution of the first level is smoothened with increasing the disorder. In Fig.~\ref{first_level}B we show the evolution of the distribution with increasing number the impurities from $N=1$ to $N=2$ and $N=7$. We clearly see how the distribution becomes more broad. In Fig.~\ref{first_level}C it is seen that the distribution becomes more or less similar for larger number of impurities, $N=5, 6, 7$

Microwave experiments~\cite{microwave} are an established way to measure the eigenfrequencies for the wave equation, system given by Eqs.~(\ref{laplas}), (\ref{bounlaplas}), in various geometries. Therefore we can propose studying microwave radiation in the cavity as a direct way of measuring the computed distributions. The boundary condition is imposed by placing highly reflective material. One can think of a simple model: a big cavity with randomly placed disks. While changing the positions of the disks and measuring the first few eigenfrequencies, the distribution for the average density and the spacing between nearest neighbors can be recovered. In this system, the lowest-lying eigenvalue statistics must be the same at the one we described above.

\section{Phonon-assisted Tien-Gordon effect}
\label{phonTG}

In this Section, we investigate how the low-lying eigenfrequencies of the oscillating graphene membrane can be experimentally assessed by means of conductance measurement.

The electrons traversing graphene are coupled to the phonon motion of the membrane as~\cite{VonOppen}
\be V_{el-ph}= \left( \begin{array}{cc}
g_1 (u_{xx}+u_{yy}) & g_2 f^*(u_{ij}) \\
g_2 f(u_{ij}) & g_1 (u_{xx}+u_{yy}) \end{array} \right) \ , \ee
where $g_1$ is the deformation potential, $g_1\sim 20-30 eV$, affected by screening;
$g_2$ is the coupling to pseudomagnetic fields, $g_2 \approx 1.5~eV$,
and $f(u_{ij͔}) = 2u_{xy} + i(u_{xx} − u_{yy})$.
For the second-order process involving out-of-plane phonons one obtains
\be V_{el-ph} \sim  -\frac{q_1q_2}{2}\phi_{q_1}\phi_{q_2} \left( \begin{array}{cc}
g_1^{sc} & -i g_2  \ , \\
i g_2  & g_1^{sc} \end{array} \right) \label{elphon}\ee
where the similarity sign shows that we do not take into account the angle of the
scattering between different phonon modes.
We can make the following estimations of the parameters in Eq. (\ref{elphon}), $q_n\sim n/L$, $g_1\sim 10 eV$ (including some screening).
The amplitude of the oscillations $\phi$ can be estimated in several different  ways:
as the amplitude of zero-point motion, $\phi=({\hbar}/(2M\omega))^{1/2}$,
or as the amplitude of thermal fluctuations, $M\omega_t^2 \phi^2 = \kappa_B T$,
or it can be varied by changing the amplitude of the oscillations to which one pumps the energy (for example, by placing the sample on the piezosubstrate).

\subsection{Generalized Landauer formula for Tien-Gordon effect}

\begin{figure}[ht!]
\begin{center}
\includegraphics[width=0.95\linewidth]{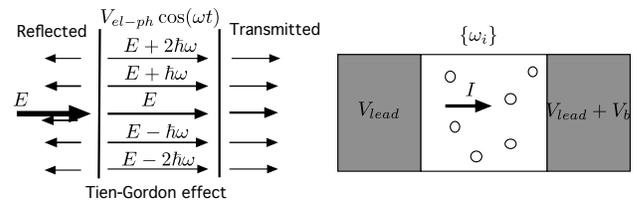}
\caption{\label{tien_gordon}
Left: While propagating through the region with potential oscillating with the frequency $\omega$  an electron with energy $E$ can scatter to the energies $E+n\hbar\omega$.
Right: Schematic representation of the current flowing from left to the right through the graphene membrane oscillating with the frequencies $\{\omega_i\}$. The Tien-Gordon effect happens due to the electron-phonon coupling.}
\end{center}
\end{figure}

Let us consider electrons in graphene subject to oscillating potential.
Electrons can emit or absorb quanta of energy corresponding to the oscillations due to the electron-phonon coupling.
Hence an electron with the energy $E$ which is coming to the region with a harmonically oscillating potential with frequency $\omega$ can be transmitted and reflected to the energies $E\pm n \hbar \omega$. A similar effect of the transmission of the electron from the energy $E$ to the energy $E+n\hbar\omega$ due to microwave irradiation of the sample is called the Tien-Gordon effect~\cite{Tie63}. Note however that there is an important difference, since in the Tien-Gordon effect the bias voltage oscillates. Our system is formally equivalent to the conductor with the oscillating barrier between the leads. Electrons in the leads are in thermal equilibrium described by the distribution functions $f_{L}(E)$ and $f_{R}(E+eV_b)$, with
$V_b$ being the difference in the potentials between two leads.

The electron propagating from the left to the right can be transmitted from the energy $E_1$ to the energy $E_{2}$ or reflected to the energy $E_2$.
We denote the corresponding transmission and reflection coefficients from as $T_{E_1\rightarrow E_2}$ and $R_{E_1\rightarrow E_2}$.
For the electron propagating from the right to the left the similar coefficients are denoted as $\widetilde {T}_{E_1\rightarrow E_2}$ and $\widetilde{R}_{E_1\rightarrow E_2}$.
Transmission and reflection probabilities are determined from the solution of the scattering problem similarly to Ref.~\onlinecite{Zeb08} (see Appendix \ref{AppendixTG}). The total transmission from the left to the right is determined by the following expression:
\be
\begin{split}
\sum_q \int dE &\left(f_L(E) - \sum_i R_{E+i \hbar \omega \rightarrow E}^{(q)} f_{L}(E+i \hbar \omega) - \nonumber \right. \\
&\left. -\sum_i f_R(E+i\hbar\omega) \widetilde{T}_{E+i\hbar\omega\rightarrow E}^{(q)} \right) \ , \label{transmission}
\end{split}
\ee
where $q$ is the transverse momentum. The first term represents electrons moving to the right.
The second and the third terms represent reflected electrons moving back to the left and electrons transmitted from the right, respectively.

The change of integration variable gives:
\bea \int dE \sum_i R_{E+i \hbar \omega\rightarrow E} f_{L}(E+i \hbar \omega) =
\nonumber\\
= \int dE \sum_i R_{E\rightarrow E-i\hbar \omega} f_{L}(E).\eea
For the case of the transmission to different energies each wave function should be normalized by the flux in the direction of propagation, namely,
\be \Psi^\dagger \hat{j} \Psi = 1 , \ee
where $\Psi$ is wave function and $\hat{j}$ is current operator.
The transmission and reflection probabilities defined in such basis obey generalized unitarity condition which is the conservation of the flux during the scattering process from the mode with energy $E$ to the modes $E+i\hbar\omega$ (see also Appendix~\ref{AppendixTG}, Eq.~\ref{gen_unitarity}):
\be \sum_i R_{E\rightarrow E+i\hbar \omega} +\sum_i T_{E\rightarrow E+i\hbar \omega} = 1.\ee
Hence we can simplify the two first terms in the expression~(\ref{transmission}):
\be \sum_q \int dE \left(\sum_i T_{E\rightarrow E+i\hbar\omega}^{(q)} f_{L}(E) - \sum_i \widetilde{T}_{E\rightarrow E+i\hbar\omega}^{(q)}f_R(E) \ . \right)\label{transmiss2}\ee
We introduce the coefficients of total transmission from an energy $E$ as
\bea
T_{LR}^{E}=\sum_q \sum_i T_{E\rightarrow E+i\hbar\omega}^{(q)}, \
T_{RL}^{E}=\sum_q \sum_i \widetilde{T}_{E\rightarrow E+i\hbar\omega}^{(q)}
\eea
and get the expression of the generalized Landauer formula~\cite{Wag99,San11}
\be \int dE \left( T_{LR}^{E}f_L(E) - T_{RL}^E f_R(E) \right) \ .\label{genLandauer}\ee

\subsection{Transmission through the oscillating graphene membrane}

\begin{figure}[tb]
\centerline{\includegraphics[width=0.95\linewidth]{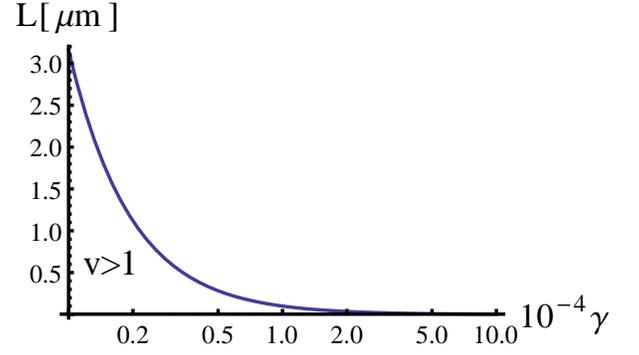}}
\caption{\label{gammaL}
The region of the parameter space of the length of the membrane and the fabrication tension $(L,\gamma)$ where the electron motion is strongly coupled to the oscillations of the membrane. The region below the curve corresponds to strong coupling.}
\end{figure}

\begin{figure}[ht!]
\begin{center}
\label{eps_v}
A\includegraphics[width=0.95\linewidth]{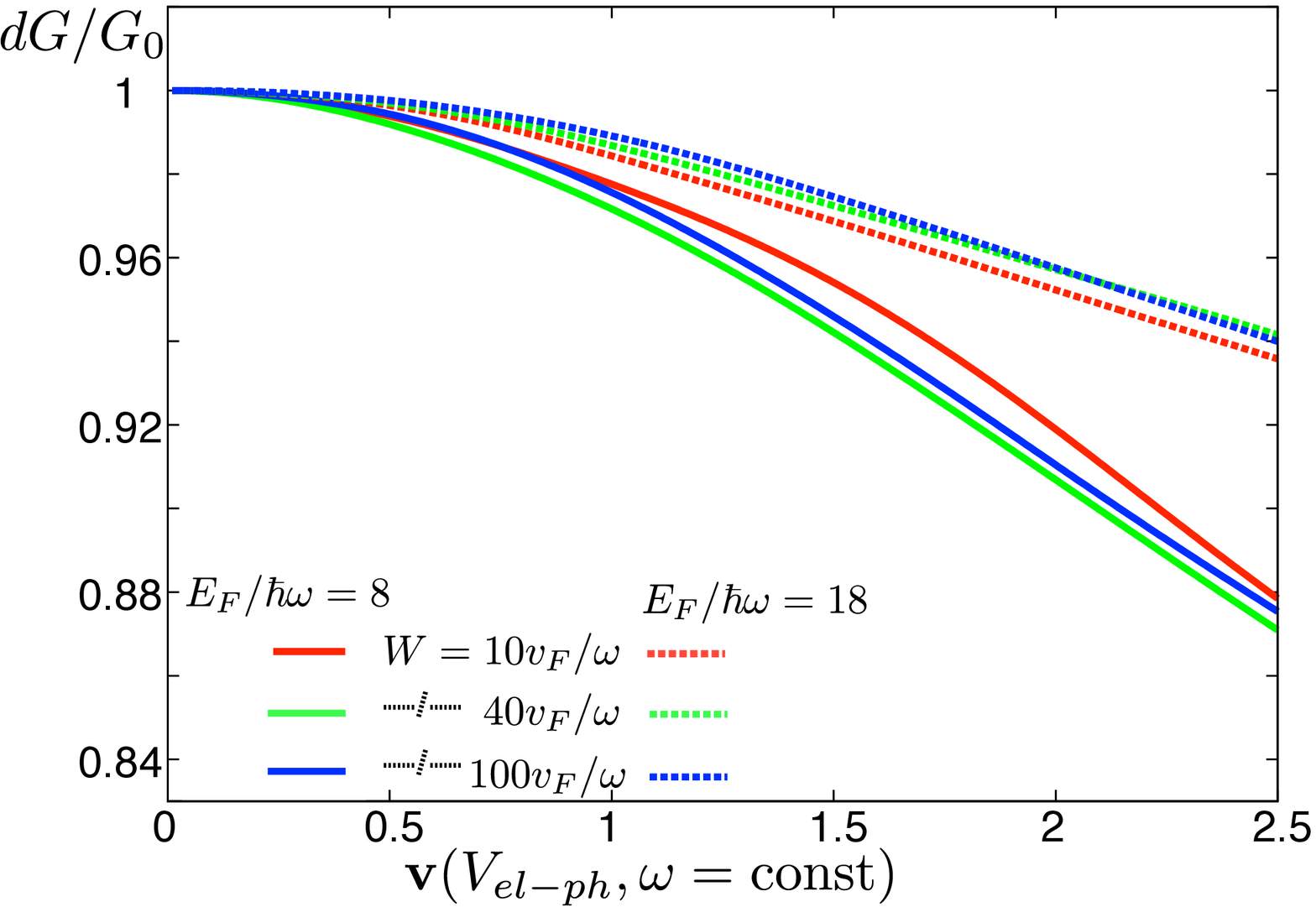}
B\includegraphics[width=0.95\linewidth]{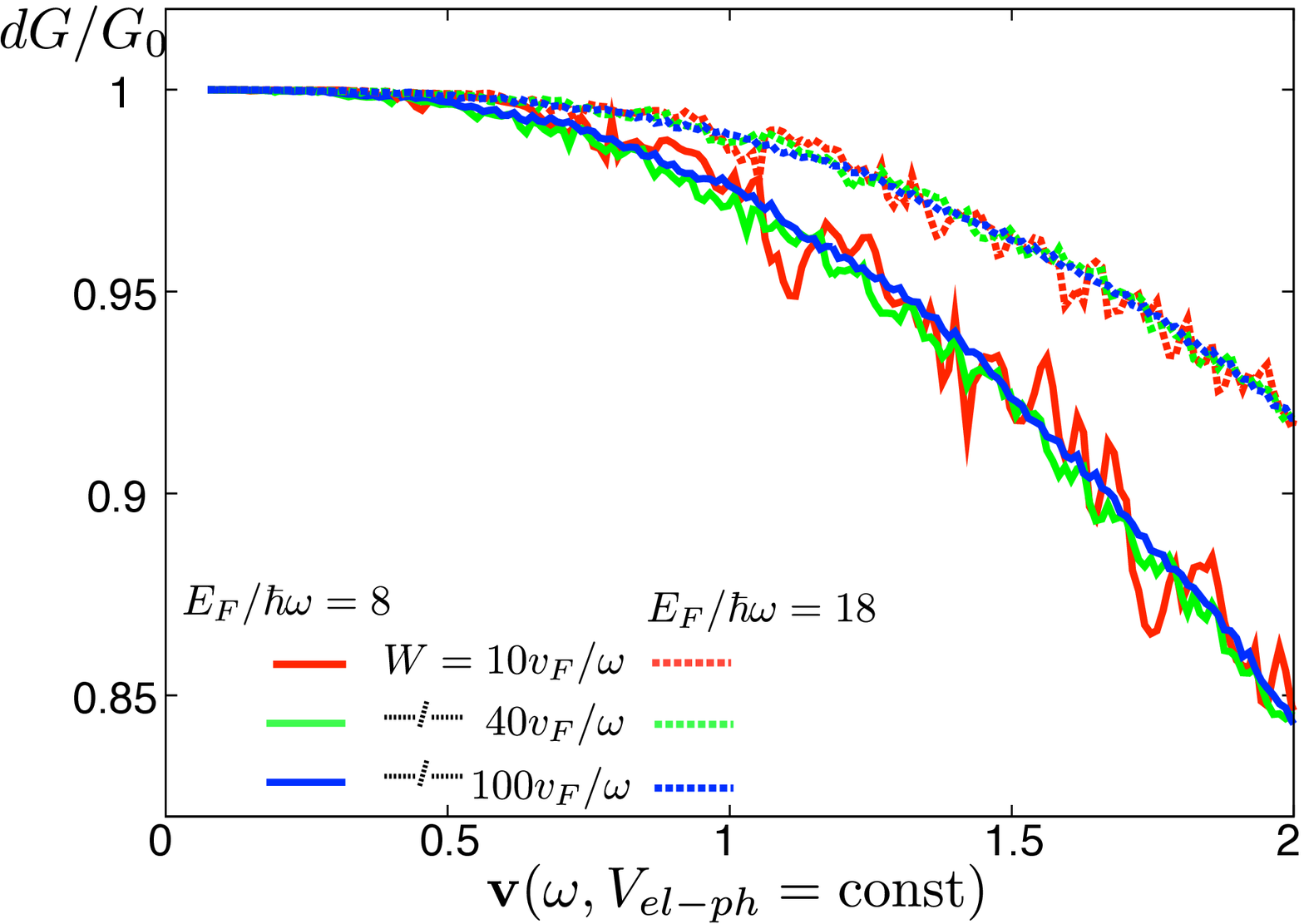}
\caption{
Dependence of the relative change of conductance $dG/G$ on the strength of the oscillations $\mathbf{v}=V_{el-ph}/(\hbar\omega)$ for fixed Fermi energy $E_F$ (gate voltage).
A: $d G/ G$ with changing electron-phonon coupling keeping  frequency of the oscillations  fixed for two Fermi energies $E_F/\hbar\omega=8$ (solid lines), $E_F/\hbar\omega=18$ (dashed lines) and three different widths  $W=10v_F/\omega,~40 v_F/\omega,~100 v_F/\omega$ and for the length $L=10 v_F/\omega$. Larger width corresponds to more $k_y$ harmonics taken into account that is why the curves with larger $W$ are more smooth.
B: $d G/ G$ for two Fermi energies $E_F/V_{el-ph}=8$,
 $E_F/V_{el-ph}=18$,
fixed electron-phonon coupling, changing the frequency.
The relative change of conductivity increases with increasing the strength of the oscillations.
The change of the frequency influences on the conductivity more than the change of the electron-phonon coupling even though the dimmensionless parameter of the strength of the oscillations is the same. It happens as we need to keep the values of the gate voltage and geometrical size in real parameters fixed.}
\end{center}
\end{figure}

\begin{figure}[ht!]
\begin{center}
\includegraphics[width=0.95\linewidth]{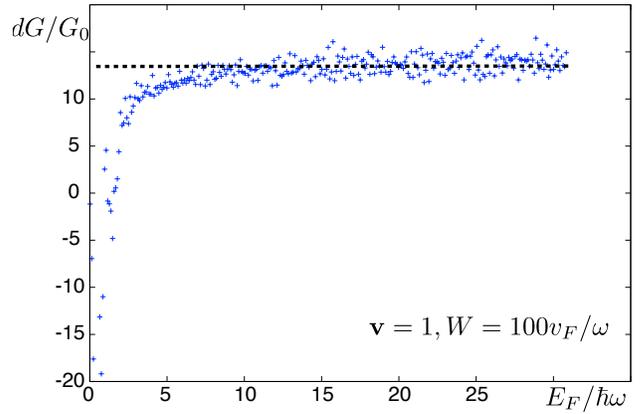}
\caption{\label{eps_vC}The absolute value of the of the conductivity correction at fixed $\mathbf{v}$ is approximately the same for different gate voltages.
The large negative values of the change of the conductivity at low gate voltages are due to Klein tunneling discussed in subsection~(\ref{subKlein}).
}
\end{center}
\end{figure}

\begin{figure}[ht!]
\centerline{\includegraphics[width=0.95\linewidth]{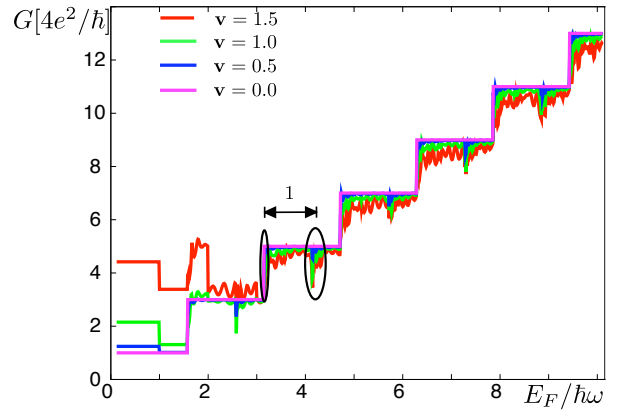}}
\caption{\label{eps_v_steps}
Dependence of the conductivity on the gate voltage for a different values of the strength of the oscillations $\mathbf{v}$ for a narrow graphene strip, $W=2 v_F/\omega$, $L=10 v_F/\omega$.
For a narrow strip the conductance is quantized. Each step of the conductance at $V_s$ is multiplied at the values of the $V_s+\hbar \omega$ as we expect.
}
\end{figure}

In further treatment, to get the conductivity of graphene with an oscillating barrier, we consider the rectangular graphene sheet. First, we get the general dependence of the conductivity on the strength of the oscillations which is determined by the influence of the substrate and the applied gate voltage (the Fermi energy in terms of the density of the electrons on the graphene flake is $E_F=\hbar v_F \sqrt{\pi n}$).
This dependence is not crucially sensitive on the shape of the graphene sheet. Second, we connect the conductivity of oscillating graphene to the distribution of the phonon frequencies (which determine the frequency of the oscillations in our setup). For generic chaotic systems such as quantum billiards it is well known that the geometric shape should not influence the distribution of the eigenfrequencies after averaging over the ensemble (in our case, over disorder configuration).

As we have shown in Eq.~(\ref{elphon}), the oscillations of the graphene membrane enter the Dirac equation which describes electrons in graphene as oscillating electrostatic potential as well as pseudovector potential. Typically, the effect of the electrostatic potential is more significant\cite{Med10}, and therefore we disregard below the effect of the pseudomagnetic fields. The electron Hamiltonian thus takes the form
\be H= - i\hbar v_F {\mathbf \sigma \cdot \nabla} + V_0 + V_{e-ph} \cos (\omega t).\ee
This problem was considered before in a number of studies: to investigate the direct analog of the Tien-Gordon effect in graphene with an oscillating lead~\cite{Tra07}, to look at the properties of transmission amplitudes depending on the angle of the incoming wave~\cite{Zeb08}, and to implement quantum pumping via evenescent modes in graphene~\cite{San11}. In our work we follow closely Ref.~\onlinecite{Zeb08} solving numerically the system of equations for transmission and reflection coefficients (as discussed in Appendix~\ref{AppendixTG}), but we go beyond this work by summing up the transmission coefficients over the different transversal momenta $q$ to obtain the conductance of the system.
Our final goal is to relate the spectrum of phonon frequencies to the features of Tien-Gordon effect.

The Tien-Gordon effect for graphene is characterized by the dimensionless parameter which is the ratio of the electron-phonon coupling $V_{e-ph}$ to the energy of the oscillations of the phonons $\hbar\omega_n$,
\be \mathbf{v_n}=\frac{V_{e-ph}}{\hbar\omega_n}.\ee

Based on the expression for electron-phonon coupling~(\ref{elphon})
we can estimate $\mathbf{v_0}$,
\be \mathbf{v_0} \sim  0.03 \cdot \frac{g_1\text{[eV]} \phi^2\text{[nm]}}{\sqrt{\gamma}L\text{[}\mu\text{m]}},\ee
where we took into account the expression for the smallest frequency of the oscillating circular membrane $\lambda_0\approx 2.4/L$.
For the case of the thermally driven oscillations with the amplitude of the oscillations $\phi^2\sim\kappa_B T/M\omega^2$ we get
\be \mathbf{v}_0 \sim  3 \cdot 10^{-11}\cdot\frac{g_1\text{[eV]} T }{\gamma^{3/2}L\text{[}\mu\text{m]}}
\label{vnum}\ee
with $\gamma$ characterizing a tension of the membrane, $\gamma=\Delta L / L$.
For example $v\sim 1$ for the realistic parameters of graphene $L=300$ nm with the tension $\gamma=10^{-4}$ with the coupling potential $g_1\sim10$ eV at the temperature $T=300$ K~\cite{foot1}. The parameter of the strength of the oscillations can be substantially modified by the tension and the size of the membrane, Fig.~\ref{gammaL}. The tension depends on the fabrication process. While increasing the tension the influence of the oscillations of the membrane on the conductivity decreases, $\mathbf{v}$ decreases. With increasing the size of the membrane $\mathbf{v}$ decreases as well. Decreasing temperature reduces the influence of the oscillations on the conductivity as the coupling is proportional to temperature. We would like to note that the higher oscillations of the membrane are less coupled to the electron motion, namely
\be \frac{\mathbf{v}_n}{\mathbf{v}_0}\sim n^{-1/2}\label{less_coupled}\ee

We determine the influence of the oscillations of the potential on the conductivity using the  generalized Landauer formula~(\ref{genLandauer}). For low temperatures, $\kappa T \ll V_{bias}$, $\kappa T \ll \hbar\omega$, we can approximate the distribution function in the leads by the constant, then the current through the system is determined by
\be I(E_0,\mathbf{v}) = V_{bias} T_{LR}^{E_0}(\mathbf{v}) \ . \ee
The transmission coefficient is already a sum over transverse momenta and over all final energies $E_0 + i \hbar \omega$ which scatter from the energy $E_0$. Here we also took into account the inversion symmetry of the system setting $T_{LR}^E=T_{RL}^E$.

For weak oscillations, $\mathbf{v}\ll1$, we can make an analytic estimation of the effect (see also Refs.~\onlinecite{Zeb08}, \onlinecite{San11}). The expression for the transmission probabilities bringing an electron from the energy $E_0$  to $E_{0}\pm\hbar\omega$ is $T_{\pm1}^0(E_0) \sim \mathbf{v}^2 T_0^0(E_0) T_0^0(E_0\pm\hbar \omega)$ (here and below we use the simplified notations for the indices of the transmission coefficients, namely, denote the energy by numbers which are $i$ in the expression $E_0+i\hbar\omega$). We can disregard all other probabilities as they are smaller in the parameter $\mathbf{v}$. The current becomes then
\be
\label{lowest}
I(E_0,\mathbf{v}) \sim I_0(E_0)(1 + \mathbf{v}^2(T_{LR}^0(E_0+\hbar\omega)+T_{LR}^0(E_0-\hbar\omega)) )\ ,
\ee
where $I_0$ is the current in the absence of the oscillations.

For moderate values of $\mathbf{v}\sim1$ we solve the scattering problem numerically and then determine the conductivity by means of the generalized Landauer formula~(\ref{genLandauer}).
In Fig.~\ref{eps_v} we present the dependence of the conductivity on the strength of the electron phonon coupling $\mathbf{v}$ for different energies on the graphene flake.
We notice that if the strength of the oscillations grows, the condictivity decreases and becomes independent on the width of the graphene sheet for large values of $\mathbf{v}$. The absolute value of the drop of the conductivity is approximately similar for different values of gate voltage on graphene.

For narrow graphene strips the conductance is quantized with increasing the gate voltage. Due to the oscillations the steps in conductance at gate voltages $V_{step}$ are reflected at the gate voltages $V_{step}+\hbar\omega$, see Fig.~\ref{eps_v_steps}.

\begin{figure}[ht!]
A\includegraphics[width=0.95\linewidth]{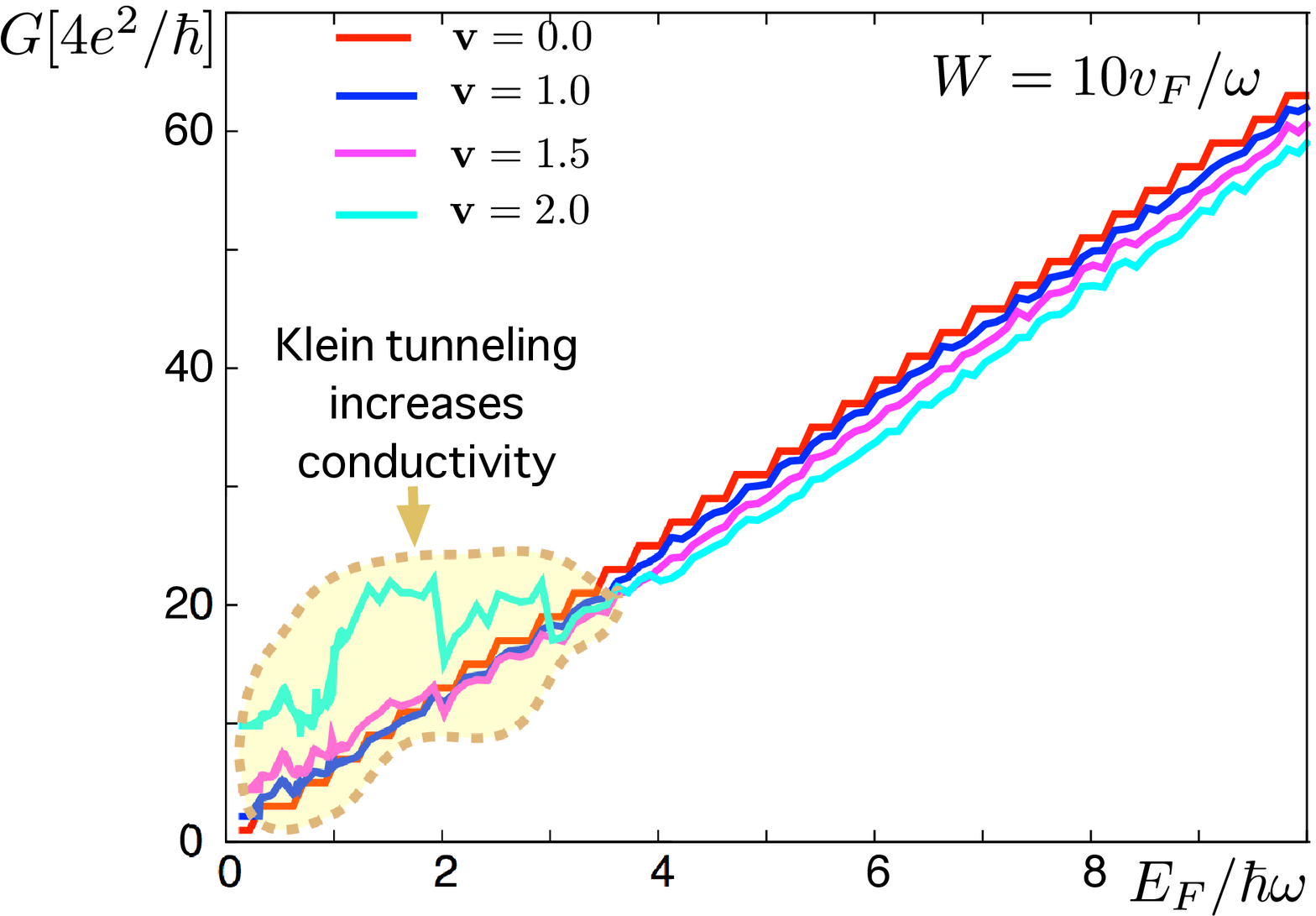}
B\includegraphics[width=0.95\linewidth]{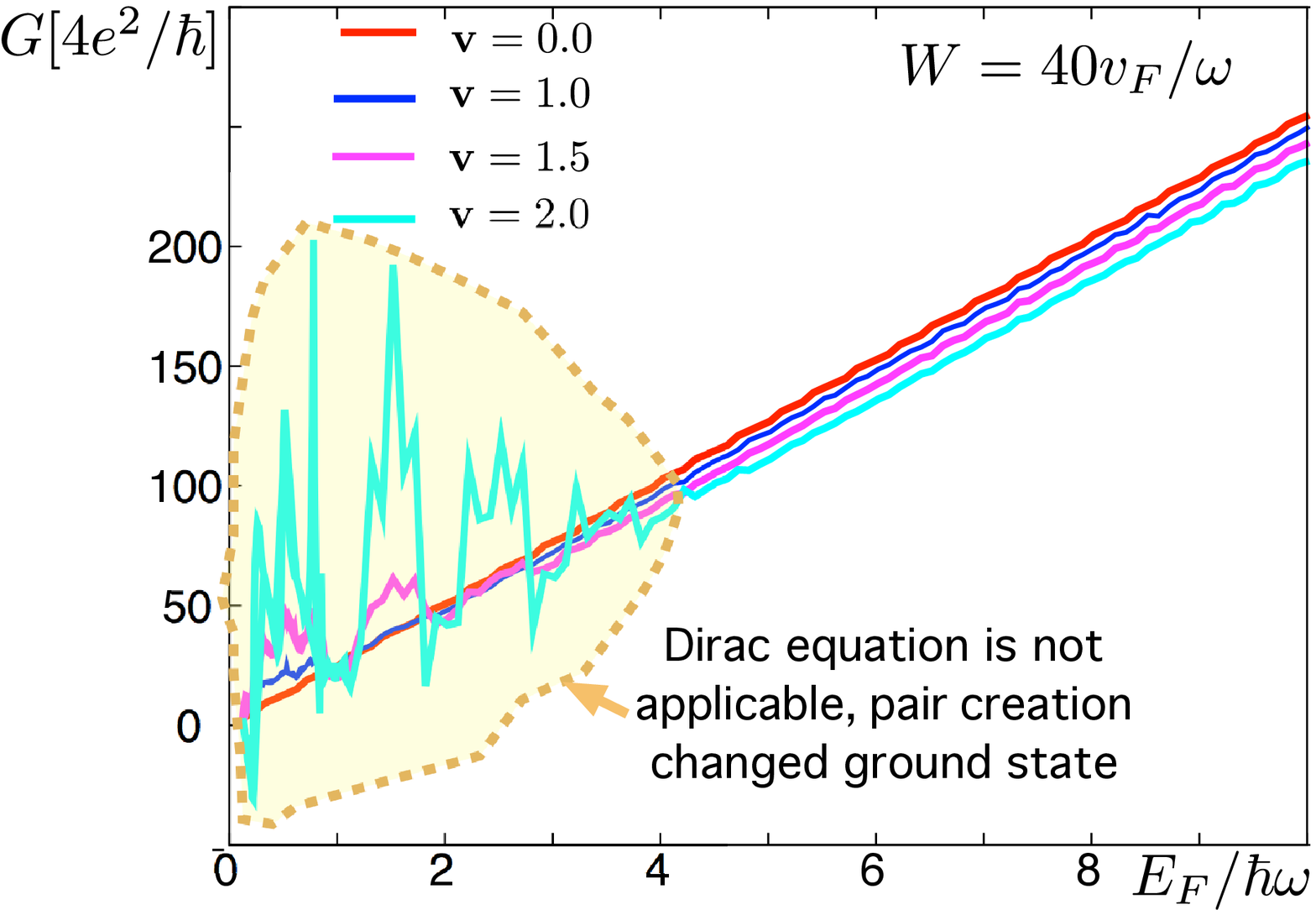}
\caption{\label{v_eps}
Dependence of the conductivity on the gate voltage for different values of the strength of the oscillations $\mathbf{v}$ for a wide graphene strip, $W=10v_F/\omega$ and $W=40v_F/\omega$.
The dependence is smooth. The conductivity slightly decreases (about 10 percent) at larger values of the gate voltage (See also Fig.~\ref{eps_vC}).
Pair creation is important at low gate voltages where conductivity significantly increases. Here the one-particle approximation with Dirac equation does not work and more precise consideration using quantum field theory should be applied.
}
\end{figure}

\subsection{Klein tunneling}
\label{subKlein}

\begin{figure}[tb]
\begin{center}
A\includegraphics[width=0.95\linewidth]{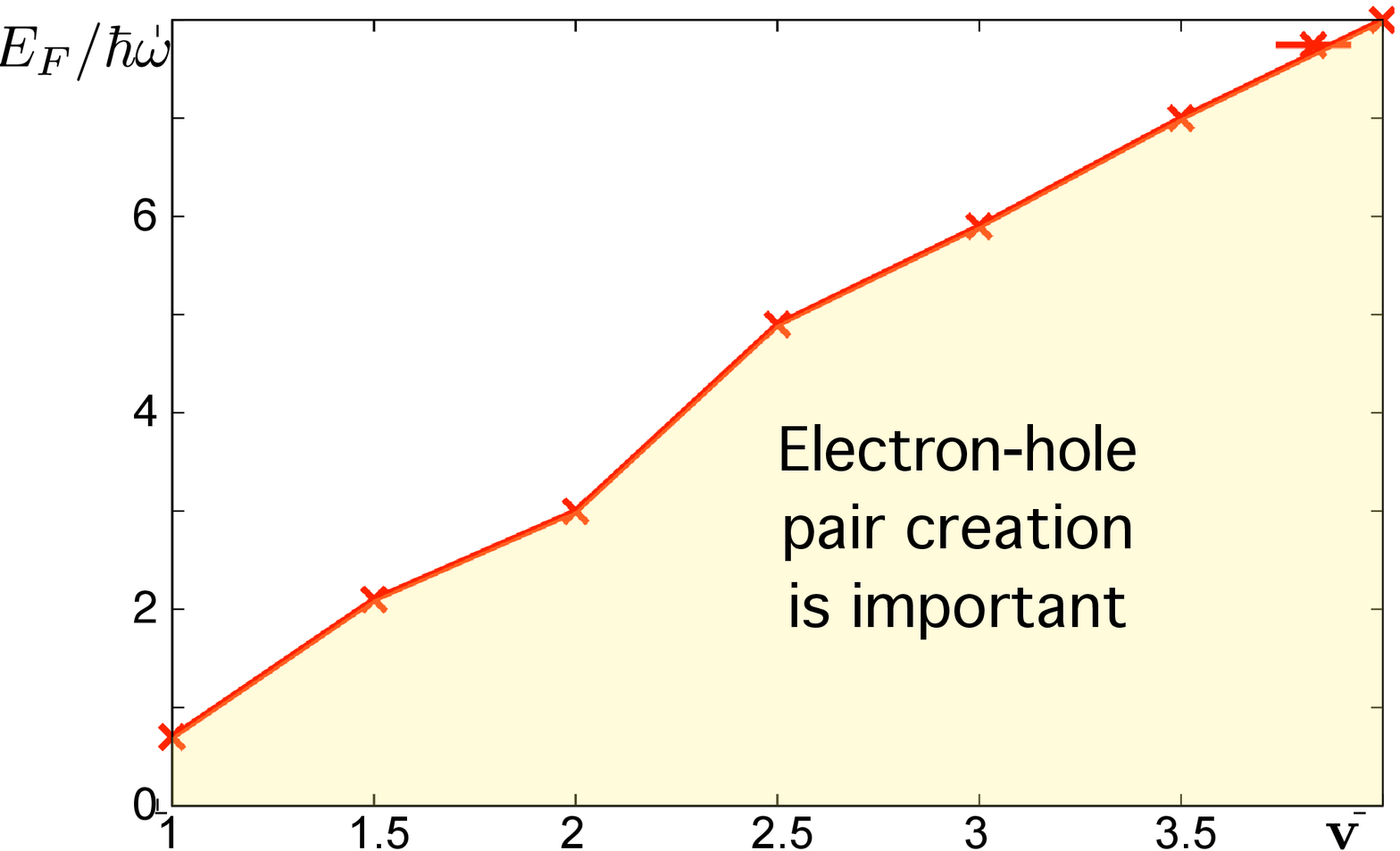}
B\includegraphics[width=0.95\linewidth]{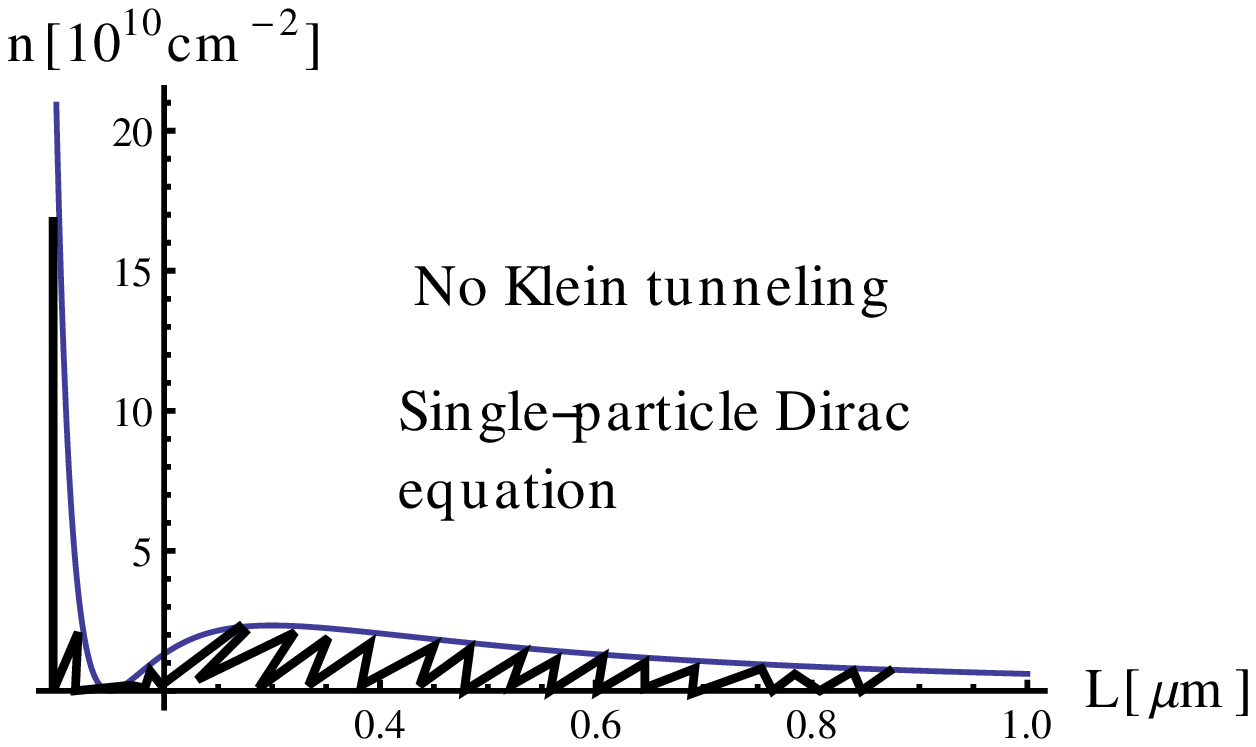}
\caption{\label{klein}
A:The area of parameters $(\epsilon,\mathbf{v})$ where the Klein tunnelling is important.
B: The region (density, size of the graphene sheet) where the pair creation is important  obtained from the curve at Fig.A recalculating from the dimensionless units to the physics units
($T=300$ K, $g_1=12$ eV, $\gamma=10^{-4}$).}
\end{center}
\end{figure}

The dependence of the conductivity on the gate voltage at high values of $\mathbf{v}$ ($\mathbf{v}>1$) and low gate voltages exhibits unexpected growth which reveals the existence of pair creation, see Fig.~\ref{v_eps}, as we explain below.

Graphene is a condensed matter material where Klein paradox can be measured directly~\cite{Katsnelson,KleinExp}.
Klein paradox arises in scattering of Dirac electrons on a high enough potential step: The tunneling probability is negative, and the reflection probability is greater then one.
Another closely related counterintuitive result is that the transmission probability remains finite for arbitrary high and long barriers and does not depend on the length of the barrier.
Both results come from the solution of the Dirac equation.
It is a consequence of one-particle treatment of Dirac fermions.
A fully consistent consideration of the problem requires the quantum field theory approach where many-particle states are taken into account.
It was done for example in Refs.~\onlinecite{Nis69,Finn81} and discussed in the context of graphene in Ref.~\onlinecite{All08}.

Klein tunneling can be represented as tunneling with a creation of an electron-hole pair. When a particle-hole pair is created, the particle can be reflected
backwards by the potential, which gives a positive contribution to the
reflection probability and can make it larger than one. The hole moves
in the same direction as the transmitted particle would move, thus
contributing a negative term to the transmission
probability~\cite{Finn81}. In classical terms the motion of the hole
is the same as motion of the particle in the opposite direction.
The above picture stems from the single-particle (quantum-mechanical)
understanding of the process. In fact, the particle-hole creation also
changes the ground state of the system and introduces vacuum
polarization. When these effects become significant, the calculation
of the transport cannot be done in the simple way as above as now the
potential itself in turn depends on the density of particles and
holes.

The equilibration of the two continua (the electron and the hole one) happens very fast, so that it is difficult to observe the pair-creation in a p-n junction directly as a characteristic time is very short.
In a time-dependent potential the pair creation occurs as well, as it is proven for the general case of slowly varying time-dependent Dirac equation~\cite{Pic05}.
An attempt to consider the time-dependent Dirac equation in the context of graphene was done in Ref.~\onlinecite{Sav11}, however, the authors did not take into account possible transmission to different energies.

%The main result of the paradox still remains in force for the time-dependent potential: The reflection probability from a high potential step can be greater than one.

Obviously  for low values of the gate voltage we encounter Klein paradox for the time-dependent potential.
Indeed, an incident particle with the energy $E$ scatters to the energies $E+n \hbar \omega$. For high strength of the oscillations $\mathbf{v}$ the scattering to the states with large $n$ can be strong. It imitates the scattering on the potential step, with $n \hbar \omega$ playing the role of the height of the potential barrier. If the energy $|E-n\hbar\omega|>E$ then the pair creation is allowed. The probability of the process becomes appreciable for sufficiently strong oscillations, $\mathbf{v}>1$. We identify the region of the parameters $(V_g, \mathbf{v})$ where the conductivity achieved from the scattering problem for the Dirac equation shows increased conductivity due to Klein paradox, Fig.~\ref{klein}.
The increase of the transmission is a measurable effect, described still by one-particle Dirac equation for the moderate values of the tranmission increase.
In the regime where transmission diverges, our results, based on single-particle Dirac equation, are not valid, and
more precise quantum field theoretical treatment is needed in order to obtain correct value of the conductivity due to the pair-creation mechanism. Nevertheless, we can claim that the conductivity at small Fermi energies increases at low values of the gate voltage and high values of the amplitude of the oscillations.

Thus, the dependence of conductivity of the gate voltage presented at Fig.~\ref{v_eps} is correct at large enough gate voltages (as it follows from Fig.~\ref{klein}A), where the single particle approach to determine the conductivity is valid, the transmission is not subject to many particle process which involve the pair creation.
We also note that the plots representing the dependence of the conductivity on the strength of the oscillations $\mathbf{v}$, Fig.~\ref{eps_v}, are made for the gate voltages large enough that the pair-creation process in not important.

The increase of the conductivity due to pair creation can be measured for small graphene samples where the coupling strength is large enough. Namely, we can approximately connect the region on the diagram~Fig.~\ref{klein}A to the density of graphene $n$ and the length of the membrane
using the relation for the Fermi-energy $E_F=\hbar v_F \sqrt{\pi n}$, frequency $\omega\sim 2.4\sqrt{\tfrac{\delta}{\rho_0}}\tfrac{1}{L}$ and the coupling strength given by~Eq.~(\ref{vnum}).
After these substitutions the region where the Klein tunneling can be significant is:
\be -1.6+\frac{2\cdot10^{-7}}{\gamma^{3/2}L\text{[$\mu$m]}} +\frac{0.02 L\text{[$\mu$m]}\sqrt{n\text{[$10^{10}$cm$^{-2}$]}}}{\sqrt{\gamma}} >0,
\label{parameters}
\ee
see Fig.~\ref{klein}B.

\subsection{Tien-Gordon effect for several modes}

\begin{figure}[ht!]
\begin{center}
A\includegraphics[width=0.9\linewidth]{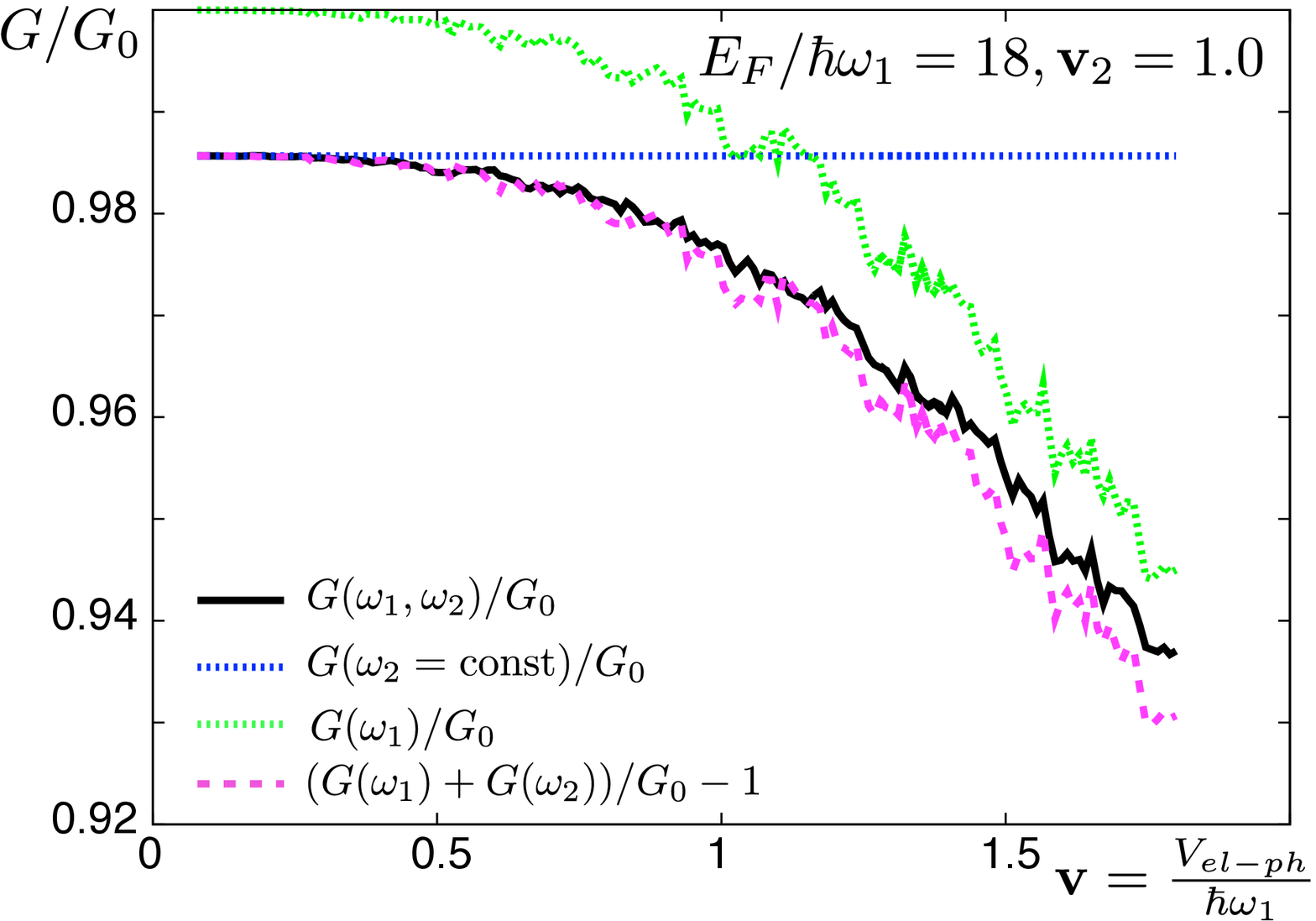}
B\includegraphics[width=0.9\linewidth]{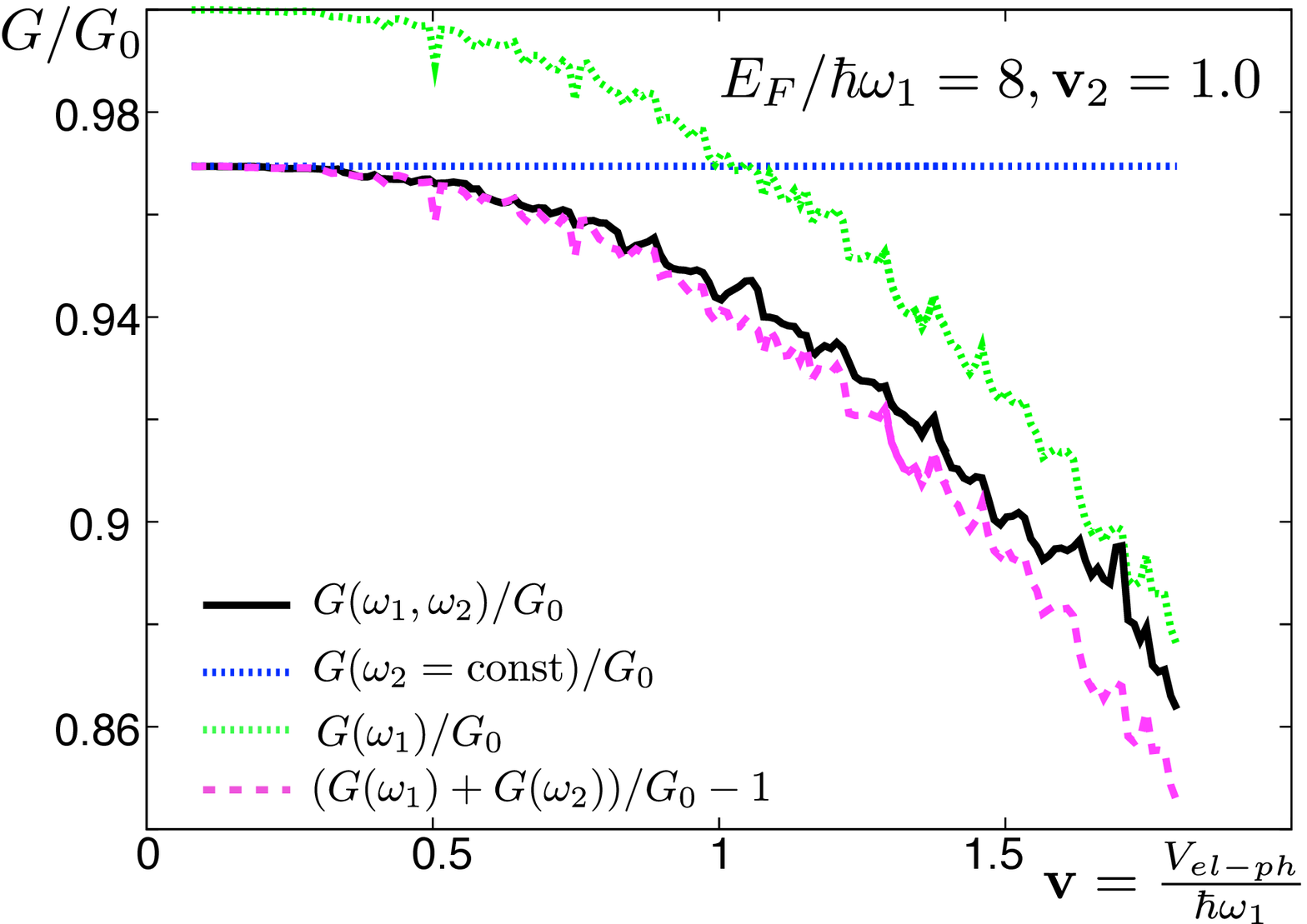}
C\includegraphics[width=0.9\linewidth]{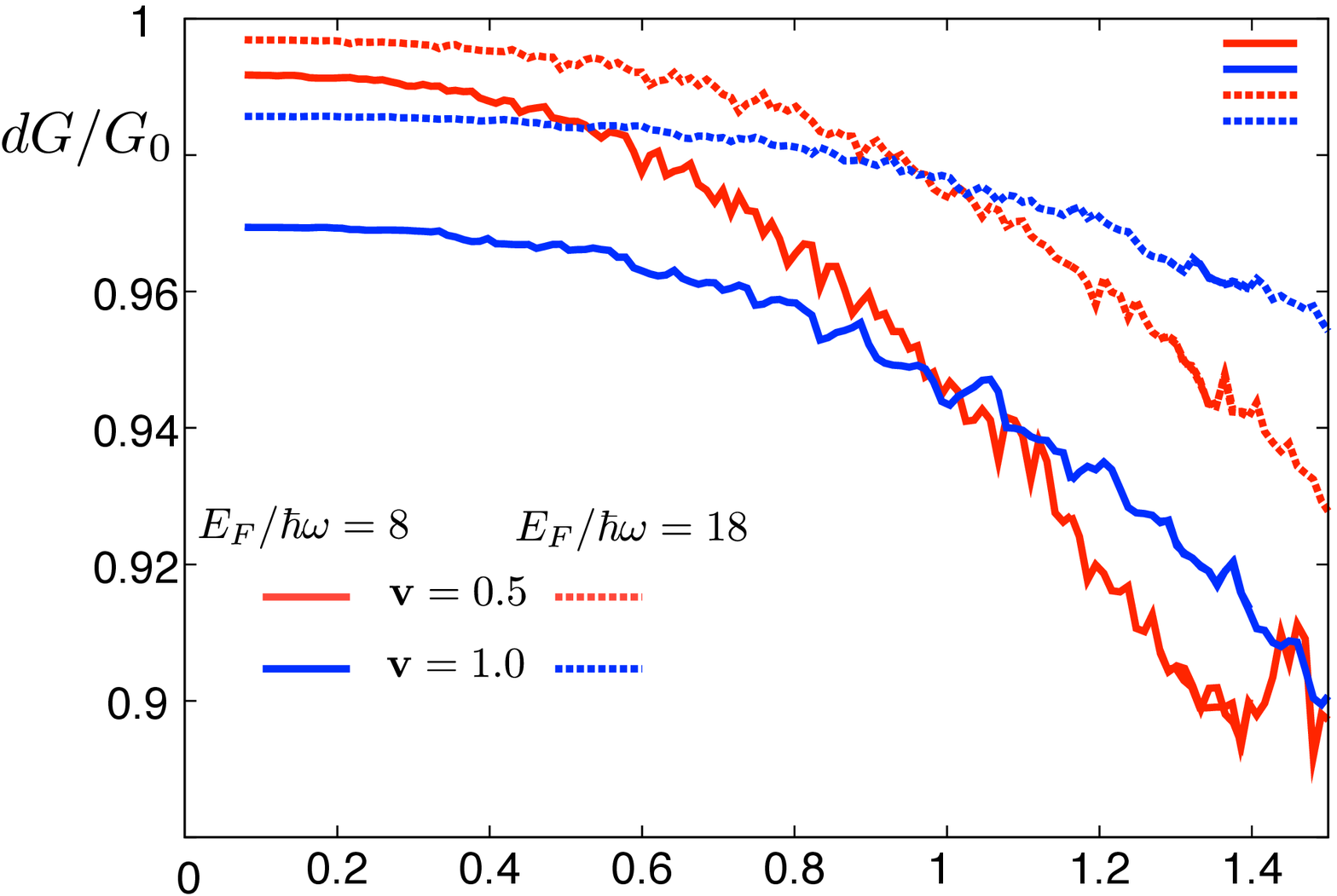}
\caption{
\label{two_modes}
The conductivity for the case of two modes of the oscillating potential. The Fermi energy is fixed.
We assume that the modes are characterized by the same coupling to the phonons $V_{el-ph}$ and different frequencies (the coupling strength depends on the phonon frequency, but for the illustration of the interaction of two frequencies this dependence is not  important). Hence the dependence on $\mathbf{v}$ translates directly to the dependence on the frequency as $1/\omega$.
A,B,C:The conductivity dependence on frequency $\omega_1$
at fixed frequency $\omega_2$ is shown. We show $G(\mathbf{v}_1,\mathbf{v}_2=const)/G_0$ and for comparison $G(\mathbf{v}_1)/G_0$ and $G(\mathbf{v}_2)/G_0$ as well as $(G(\mathbf{v}_1)+G(\mathbf{v}_2))/G_0-1$. We see that the effect of two frequencies is almost additive, A,B. Fig. C presents $G(v)$ for different sets of parameters: at smaller energies the correction to conductivity is more significant.
%Let us notice some wiggles in the dependence of the conductivity: they are present here due to finite size effects, namely due to the fact that we do not take into account the evansent modes in computing the conductivity. They are not present in the plots with only one oscillating frequency because for the case of two oscillating frequency we need to rescale the the energy of the graphene and the size of graphene in dimensionless units to keep fixed the real energy and the size in the units set by the fixed frequency [or finite size? to be compared with the width 40]
}
\end{center}
\end{figure}

Thus far, we only considered a monochromatic excitation of the membrane. In this Subsection, we look how the presence of several oscillation frequencies of the membrane influences the current. If several modes are essential, the largest contribution is given by the lowest modes, since for them $\mathbf{v}$ is the largest. The higher is the oscillation mode the less is the correction to the conductivity is contributes. As we show below there is no region for parameters in our problem where the interference between different frequencies is important for the conductivity, and the modes contribute independently: The influence of all modes is just the sum of individual contributions.

In this subsection we look first at the conductivity in the presence of two frequencies.
We solve numerically the scattering problem for the Hamiltonian,
\be H= - i\hbar v_F {\mathbf \sigma \cdot \nabla} + V_0 + V_{e-ph} \left(\cos (\omega_1 t)+\cos(\omega_2 t +\psi)\right).\ee
using the same method which we set up for the case of one frequency.

\begin{figure}[ht!]
\begin{center}
\includegraphics[width=0.7\linewidth,angle=270]{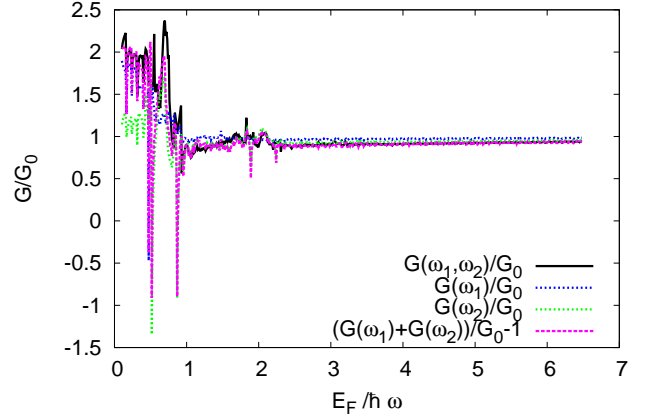}
\caption{
\label{klein_two}
Dependence of the conductivity on the Fermi energy in the case of a two-mode oscillating barrier.
We clearly see that the additivity property is preserved even for low energies where Klein tunneling plays an important role.
The deviation from the additivity occurs due to finite size of the system and lack of account of evanescent modes.
}
\end{center}
\end{figure}

For weak electron-phonon coupling, $\mathbf{v}_1, \mathbf{v}_2 \ll 1$ the influence of the two oscillations is additive in the lowest order approximation, see Eq.~(\ref{lowest}). Moreover, we see from the results of numerical treatment, displayed on Fig.~\ref{two_modes}, that this additivity property  is approximately valid even for moderate values of the oscillation strength $\mathbf{v}_i\sim 1$. Namely, the relative change of the conductivity for the oscillations with two frequencies $(G(\omega_1,\omega_2)-G_0)/G_0$, $G_0=G(\omega=0)$, equals to sum of the relative changes of the conductivity for the oscillations with one frequency $(G(\omega)-G_0)/G_0$,
\be
\frac{G(\omega_1,\omega_2)-G_0}{G_0}\approx
\frac{G(\omega_1)-G_0}{G_0}+
\frac{G(\omega_2)-G_0}{G_0} \ .
\label{rel_change}
\ee
In the region of low energies the Klein tunneling is preserved and the conductivities have the additive property, Fig.~\ref{klein_two}. The slight deviation from the additivity occurs since we do not take into account evanescent modes for determining the conductivity.

Taking into account this property, we can determine the disorder-average correction to the conductivity due to the presence of different phonon frequencies,
\be
\label{additive}
\left \langle \delta I\left(\frac{E_F}{\hbar\omega}\right) \right\rangle_{\text{dis.}} \approx \left\langle \sum_{i} \delta I(\omega_i)\right\rangle_{\text{dis.}} =\int P(\omega) \delta I(\omega)d \omega \ .
\ee

The influence of the high-frequency modes can be easily taken into account since their density of states is constant, $dN/d(\omega^2)=C$, which leads to $P_{\text{high}}(\omega)=dN/d\omega=2C\omega$, and the strength of the oscillations is small $\mathbf{v} = V_{el-ph}/\hbar\omega \ll 1$ for high frequencies, hence the relative change of the conductivity for every mode is proportional to $\mathbf{v}^2$, and for the range of high energy modes it is
\be
\label{highfr}
\left \langle \delta_{\text{h.lev.}} I\left(\frac{E_F}{\hbar\omega}\right) \right  \rangle_{\text{dis.}} \sim \int_{\omega_1}^{\omega_{max}} 2C\omega \left(\frac{V_{el-ph}}{\hbar \omega}\right)^2 d \omega
\ee
where the cut-off frequency $\omega_1$, which is the position of the first energy level of the oscillations, is given by the geometrical size of the membrane, and $\omega_{max}$ is the maximum frequency of the oscillations which is determined by microscopic properties of the membrane at zero temperature or by $\kappa_B T/\hbar$ at finite temperature (we assume that the maximum frequency is given by comparission the amplitude of zero point fluctuations and the amplitude of the thermal fluctuations, namely $M\omega_{max}\phi_{th}^2\sim\kappa_B T\sim \hbar\omega_{max}$).
We also take into account the decrease of the coupling with increasing the frequency of the oscillation, Eq.~(\ref{less_coupled}).
The constant density can be estimated from the general consideration: The density of states of the oscillating membrane according to the Weyl formula is $dN/d\lambda^2=(\gamma/\rho_0)dN/d\omega^2=(1-r^2n)L^2/2 \approx L^2/2$, where $n$ and $r$ are the number of holes and the radius.
Therefore the oscillations at high frequencies can decrease the conductivity by
\be
\left\langle \frac{\delta_{\text{h.lev.}} G}{G_0} \right\rangle_{\text{disorder}} \sim  \left( \frac{V_{el-ph}}{\hbar\omega_1} \right)^2 \ln \left( \frac{\kappa_B T}{\hbar\omega_1}\right) \ .
\ee
We can express the change of the conductivity for the real sample:
\be
 \left\langle \frac{\delta_{\text{h.lev.}} G}{G_0} \right\rangle_{\text{dis.}} \sim
0.7\cdot 10^{-21} \frac{g_1^2 \text{[eV]} T^2\text{[K]} }
                     {L^2\text{[$\mu$m]} \gamma^{3}}
                  \ln \left(\frac{3 T \text{[K]} L\text{[$\mu$m]}}
                                 {\sqrt{\gamma}} \right)
\ee
The larger is the membrane the smaller is the conductivity correction.
For large values of the electron phonon coupling $V_{el-ph}>\hbar\omega_1$ one observes a noticeable decrease of the conductivity due to first several modes of the oscillations. For example,
for the graphene membrane of the length $L=500$ nm and the tension $\gamma=0.3\cdot10^{-3}$ the correction is $\left\langle \frac{\delta_{\text{high levels}} \sigma(\epsilon)}{\sigma_0} \right\rangle_{\text{disorder}} \sim 0.015$.
Let me note that the correction is very sensitive to the stretching.

\section{Discussion} \label{discuss}

In this Article we investigated the statistics of low-energy spectrum of the graphene membrane suspended over disordered substrate and the effects of this statistics on the conductivity. We modeled the graphene sheet as a membrane attached to randomly located circular areas. The phonon frequencies in this system are given by eigenvalues of the Laplace equation with corresponding boundary conditions.

It is known that the high-energy eigenvalues in this system obey the Wigner-Dyson statistics. This is valid for a generic chaotic system, where the statistics is found as the result of the averaging over the energy with the constant density of states. It was less obvious what the statistics of low-lying eigenfrequencies is, since conventional analytical methods are not applicable to this situation. For a single sample (with a given impurity configuration) one can not define the statistics, since one can not average over energy. However, in a disordered system one can instead average over impurity configurations (positions of the attachment areas). We performed this averaging numerically and found that the statistics of the nearest level spacing between the first five-six levels obeys the Wigner-Dyson probability distribution. The important difference between high-energy and low-energy case is that the density of eigenvalues is not constant for low energies --- rather, it oscillates for the first few levels, which is a remnant of the existence of the positions of the eigenvalues of a clean circular membrane. At high energies, the density of eigenvalues saturates. Note also that the eigenvalues of a disordered membrane can never lie lower than the lowest eigenvalue of the clean membrane, which creates an offset energy in the density of states. While changing this cut-off, in physical terms changing the size of the membrane, the statistics of the levels is not changed much.

Our results can not be described by the non-linear $\sigma$-model, a standard tool to deal with disordered systems, because in our system the disorder is not strong enough. We note that whereas in the non-linear $\sigma$-model effective bosonic fields are introduced after averaging over the disorder, and these fields are confined to a sphere (compact sector) due to strong disorder, in our case of low-lying levels such fields would not be confined to the sphere. The indirect evidence for this is the oscillation in the density of levels for the first few levels.

The phonon levels could be accessed directly, for example, by coupling the graphene membrane to other degrees of freedom like microwave photons. However, these experiments presently look too involved, and therefore we investigated the effect of the phonons on electron transport through graphene. The coupling between electrons and phonons is provided by the deformation potential. The electron transport in this situation shows the phonon-assisted Tien-Gordon effect. Due to the oscillations of the membrane with the frequency $\omega$ the electron can be transmitted from the energy $E$ to the energy $E+n\hbar\omega$.
At strong electron-phonon coupling and a small membrane we find that the influence of the phonon-assisted Tien-Gordon effect on the conductivity can be of order of ten percent. If the frequency increases, the influence of the phonon oscillations becomes less significant, and thus the lowest frequencies of the phonons have the most crucial influence on the change of the conductivity. The modes affect the current independently, and thus the influence of the superposition of the modes can be approximated by the sum of the influences of an individual mode.

The Tien-Gordon effect is the most pronounced in the case of the presence of structure in the conductivity.
The structure, in our case the Fabry-Perot resonances, appears in the voltage dependence of the conductivity in the case when tunnel barriers are located at the ends of the sample. Then every resonance at the energy $E$ has its sattelites at the energies $E+n\hbar\omega$.
This can be also a way to measure the phonon frequencies. This method can be applied only to the case of the narrow samples, as for real two-dimensional sample the Fabry-Perot resonances have different positions for different transverse momenta, and summing over the transverse momentum makes the Fabry-Perot peaks and their satellites indistinguishable.

At low gate voltages and strong electron-phonon coupling the oscillation potential can create the potential step large enough to detect Klein tunneling process, namely, increase of the conductivity due to electron-hole pair creation. There is a range of parameters where the single particle Dirac equation is applicable to describe this process. When the reflection and transmission probabilities start to diverge, many-particle quantum field theoretical treatment should be performed. Unexpected increase of the conductivity at low gate voltages can be a signal of strong electron-phonon coupling. These effects lie outside the scope of this Article.

Let us note that a suspended graphene membrane is not the only situation when low-lying eigenvalues of the Laplace equation can be assessed. For instance, our treatment also applies for a system of microwave cavities with randomly positioned reflective disks inside. The distribution of photon frequencies in these cavities will have the same features as described in this Article.

\section*{Acknowledgments}

\noindent
We acknowledge the financial support of the Future and Emerging Technologies program
of the European Commission, under the FET-Open
project QNEMS (233992), of the Dutch Science Foundation NWO/FOM and of the Eurocores program EuroGraphene. M. M. thank M. \v{C}ubrovi\'c for discussions on quantum chaos.

\appendix
\section{Boundary integral method for a region with holes}
\label{BoundaryIntegral}

\begin{figure}[ht!]
\begin{center}
\includegraphics[width=0.85\linewidth]{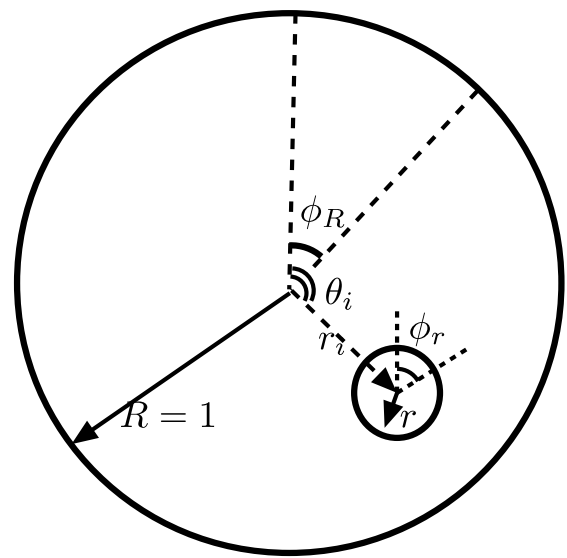}
\label{figappendix}
\caption{The parameters used in Eq.~(\ref{qf})~-~(\ref{ql}).
}
\end{center}
\end{figure}

In this Appendix, we show how to solve the eigenvalue problem for the Laplace equation:
\be \Delta \phi(q) + \lambda_n^2 \phi(q) =0, \text{ for $q$ in }\Omega/\pp\Omega\ee
with the boundary condition
\be \phi(q) =0 \text{ for $q$ in } \pp \Omega.\ee
The solution can be obtained using the boundary integral method widely implemented for the solution of this
problem for quantum billiards~\cite{Baecker}, with the only difference that we are interested in a problem
with multiple boundaries, namely, a region with holes inside it.

Let us first introduce the general formalism of the boundary integral method and then
specify the problem for the case of several boundaries.

We consider the normal derivative of the oscillating field~\cite{Baecker}
$u(s)=(\pp/\pp \mathbf{n})\phi(\mathbf{q}(s))=\mathbf{n}(s)\mathbf{\nabla} \phi(\mathbf{q}(s))$,
where $s$ is a parametrization of the boundary and $n$ is the normal consistent with the direction around the contour.
The field obeys the integral equation,
\be u(s) = \int_{\pp \Omega} Q_n(s,s')u(s') ds', \label{BIE}\ee
with the integral kernel
\be Q_n(s,s')=-2\frac{\pp}{\pp n} G_n(q(s),q(s\prime)) \ , \label{kernel}\ee
where $G_n(q(s),q(s\prime))$ is the Green's function of the corresponding Laplace equation,
\be (\Delta +\lambda_n^2)G_n(\mathbf{q},\mathbf{q}')=\delta(\mathbf{q}-\mathbf{q}') \ . \ee
In two dimensions, the Green's function of free motion is Hankel's function, $H_0^{(1)}$.
Therefore the kernel~(\ref{kernel}) can be expressed as
\be Q_n(s,s')=-\frac{i\lambda_n}{2} \cos \alpha(s,s')H_1^{(1)}(\lambda_n\tau(s,s')),\label{kernelgeneral}\ee
with $\tau(s,s')=|q(s)-q(s')|$ and
$$ \cos\alpha(s,s')=\frac{n(q)(q(s)-q(s'))}{\tau(s,s')}.$$

For numerical implementation the equation~(\ref{BIE}) is discretized, the boundary is discretized, $s_{(i)}$, and the corresponding values of $u_{(i)}=u(s_{(i)})$ are taken. The condition that the solution exists is
\be \det \left(Q_{n(i,j)}-1\right)=0. \label{discr}\ee
We vary $\lambda$ and monitor the left-hand side of the condition~(\ref{discr}), thus finding the eigenfrequencies of the Laplace equation. Let us note that numerically it is more efficient and more precise to look at the eigenvalues of $Q_{n(i,j)}-1$ then to look at the condition for the determinant.

The boundary integral method can be simply generalized for a region with holes.
Let us represent the boundary as $\pp \Omega = \pp \Omega_0 \cup \pp \Omega_1 \cup \dots\cup \pp \Omega_N$
where $\pp\Omega_0$ is outer boundary of the area and $\{\Omega_i\}$, $i = 1,\dots,N$ are holes inside of the area.
The directions of the boundaries and normal vectors to the holes are consistent with the direction of the outer boundary. %\ref{fig}.

Now the function $u(s)$ on the boundaries is represented as $\left( u_{\pp \Omega}, u_{\pp \Omega_1},\dots,u_{\pp \Omega_N} \right)$.
For further convenience of the notations we use the indices $\Omega$ and $\Omega_i$ instead of the precise notations $\pp \Omega_0$, $\pp \Omega_i$.
The system of integral equations is:
\be u_{\pp\Omega}= \int_{\pp\Omega} Q_n^{\Omega\Omega}u_{\Omega}+\sum_{i}\int_{\pp \Omega_{i}} Q_n^{\Omega \Omega_i }u_{\Omega_i} \ ,\ee
\be u_{\Omega_i}= \int_{\pp \Omega} Q_n^{\Omega_i \Omega }u_{\Omega}+\int_{\pp \Omega_{i}} Q_n^{\Omega_i \Omega_i}u_{\Omega_i}+
\sum_{j,j\ne i}\int_{\pp \Omega_{j}} Q_n^{\Omega_j \Omega_i}u_{\Omega_j}  \ , \ee
with $Q_n$ being the corresponding kernels given by general formula (\ref{kernelgeneral}).

For example we present the kernels for our model of the circular outer membrane and the holes of radius $r$ placed in the random positions inside of the membrane. We parametrize the hole position by the position of the center and the angle $(r_i,\theta_i)$. The position on the outer membrane is given by angle $\phi_R$, on the small circle by $\phi_r$. We take the radius of the outer membrane $R=1$ and for the holes $r$. All parameters are shown at~Fig.~\ref{figappendix}. By $\tau$ we denote the distance between the points for which the kernel element is given, $\tau(s,s')=|\mathbf{q}-\mathbf{q}'|$.
\begin{widetext}
\begin{itemize}
\item For $\mathbf{q}$ at the large circle and $\mathbf{q'}$ at small circle is (the negative sign originates from the circulation in another direction for the integral over the small circle)
\be \label{qf}
Q_n^{\Omega \Omega_i}(s,s')=\frac{i\lambda_n}{2}
\frac{1- r\cos(\phi_R+\phi_r)-r_i\cos(\phi_R-\theta_i) }
{\tau}
H_1^{(1)}(k\tau(s,s')),\ee

\item for $\mathbf{q}$ at the small circle and $\mathbf{q'}$ at large circle is (the negative sign originates from the direction of normal)
\be
Q_k^{\Omega_i \Omega}(s,s')=-\frac{i\lambda_n}{2}
\frac{r+r_i \cos(\phi_r+\theta_i)-\cos(\phi_r+\phi_R) }
{\tau}
H_1^{(1)}(k\tau(s,s'))\ee

\item for both $\mathbf{q}$ and $\mathbf{q'}$ at the small circle (one negative sign originates from  normal, and another one from the direction)

\be
\label{ql}
Q_k^{\Omega_i \Omega_j}(s,s')=
-\frac{i\lambda_n}{2} \frac{r_i \cos(\theta_i+\phi_{ri})-r_j\cos(\phi_{rj}+\theta_j)-r\cos(\phi_{ri}-\phi_{rj})}
{\tau}H_1^{(1)}(k\tau(s,s'))\ee
\end{itemize}
\end{widetext}

\clearpage

\section{Scattering problem for the Tien-Gordon effect in graphene}
\label{AppendixTG}

We consider scattering of electrons in graphene in the time-dependent periodic potential,
\be H= - i\hbar v_F {\mathbf \sigma \cdot \nabla} + V_0 + V \cos (\omega t),\ee
where $\mathbf{\sigma}=(\sigma_x, \sigma_y)$ are the Pauli matrices, $V_0$ is the potential in the region of propagation,
$V$ is the amplitude and $\omega$ is the frequency of the oscillating potential.
Due to the oscillating potential the incident wave with the energy $E$ scatters into waves with energies $E+n \hbar \omega$.
Our goal is to determine the transmission coefficients and to get the conductivity through the region of graphene with
the oscillating potential.

The time-dependent wave-function of the incident Dirac electron with energy $E$ is:
\be \Psi_{in}(t,x,y)=e^{ik_y y} \begin{pmatrix} 1 \\ z_{k_{in}^{(0)},k_y}\end{pmatrix} e^{ik_{in}^{(0)}x} e^{-iEt/\hbar}, \ee
with the momenta $k_y$ and $k_{in}^{(0)}$ perpendicular and along the strip respectively,
such that $k_y^2+(k_{in}^{(0)})^2=E^2$,
and the phase factor $z_{k,k_y}=(k+i k_y)/E$.
Reflected and transmitted wave functions are written as sums over the energies $E+l \hbar \omega$,
\bea \Psi_r(t,x,y)= e^{ik_y y} \sum_{l} r^{(l)}_{k_y} \begin{pmatrix} 1 \\ z_{-k_{in}^{(l)},k_y}\end{pmatrix} e^{-ik_{in}^{(l)}x} e^{-i(E+l \hbar \omega)t/\hbar},\nonumber\\
\Psi_t(t,x,y)=e^{ik_y y} \sum_{l} t^{(l)}_{k_y} \begin{pmatrix} 1 \\ z_{k_{t}^{(l)},k_y}\end{pmatrix} e^{-ik_{t}^{(l)}x} e^{-i(E+l \hbar \omega)t/\hbar} \nonumber \eea
with the momenta $k_t^{(l)}$ such as $$k_y^2+(k_t^{(l)})^2=(E+l\hbar\omega+V_b)^2.$$
The wave function in the region with the time-dependent potential has the time dependent factor $\exp(-iV/\hbar\int \cos \omega t)$ which can be expanded over the harmonics $\exp(i n \omega t)$~\cite{Tie63},
\be  \exp\left(-i\frac{V}{\hbar}\int \cos \omega t\right)=\sum_{-\infty}^{\infty}J_n\left(\frac{V}{\hbar \omega} \right)\exp(-in\omega t) \ , \ee
where $J_n$ is the Bessel function of $n$th order.
Hence the wave function in the oscillating region is expressed as:
\begin{widetext}
\be \Psi_b(t,x,y)=e^{ik_y y} \left[ \sum_{l}  e^{-i(E+  l \hbar \omega)t/\hbar} \left( A^{(l)}_{k_y} \begin{pmatrix} 1 \\ z_{k_{b}^{(l)},k_y}\end{pmatrix} e^{ik_{b}^{(l)}x}
+B^{(l)}_{k_y} \begin{pmatrix} 1 \\ z_{-k_{b}^{(l)},k_y}\end{pmatrix} e^{-ik_{b}^{(l)}x}
\right) \right]\sum_n J_n\left(\frac{V}{\hbar \omega} \right)\exp(-in\omega t)
\ee
\end{widetext}
with the momenta $k_b^{(l)}$ along the strip that obey $k_y^2+(k_b^{(l)})^2=(E+V_0+l\hbar\omega)^2$.
The conditions of the continuity of the wave function at $x=0$
$$\Psi_{in}(t,0,y)+\Psi_r(t,0,y)=\Psi_b(t,0,y),$$
and at $x=L$, $$\Psi_b(t,L,y)=\Psi_t(t,L,y),$$
give the system of linear equations for the reflection and transmission coeffients.

The obtained system of equations is infinite.
We solve the system numerically making it finite and ensure that the result does not depend on the number of the equation in the system.

Let us note that the above system of equations takes into account both propagating and evanescent modes in the region between the leads.
To determine the transmission and reflection probabilities we should take into account the
normalization by unit flux in $x$ direction for the mode of every energy,
\be N_E (1 , z^{*}_{k,k_y}) \sigma_x
 \begin{pmatrix} 1 \\ z_{k,k_y}\end{pmatrix} =1 \ ,
\label{flux} \ee
which leads us to $N_E = 2 k / E$. Note that the evanescent modes can not be normalized in this way. However, in the following we assume that the system is sufficiently far from the Dirac point, and the effect of evanescent modes on electric transport is insignificant.
Then the transmission and the reflection probabilities are
\be T_l = \frac{k_{t}^{(l)}  E}{(E+l\hbar\omega+V_b) k_{in}^{(0)} }|t_l|^2,\ee
and
\be R_l = \frac{E} {E+l\hbar\omega+V_b} |r_l|^2.\ee
The flux conservation condition~(\ref{flux}) is given by
the generalized unitarity condition (generalized as we take into account several modes):
\be \sum_l \left( T_l +R_l\right)=1.\label{gen_unitarity}\ee
We ensure this property numerically to check that the numerical size of the system of the equations is sufficient to preserve the current through the system.

The same system of equations was considered in Ref. ~\onlinecite{Zeb08}. We extend the analysis of that publication, since
at the end we determine the full transmission probability, i.e. we sum over all propagating modes. We generalize the problem to several frequencies as well.
For example for the case of two frequencies $\omega_1$ and $\omega_2$, the scattering will happen to the energies $E+l\hbar\omega_1+n\hbar\omega_2$. If for the treatment of the system with only one frequency one needs the system of the equations of the size $4N$, then for the treatment of the system with two frequencies it is $4N^2$. This growth of computational complexity prevented us from looking at the strong coupling region where both parameters $V/\hbar\omega_{1,2}$ are large.


\begin{thebibliography}{99}
\bibitem{foot1} Note that the dependence on $\gamma$ in the denominator leads to divergence at $\gamma\longrightarrow 0$, but the tension is never zero even in the absence of initial pretension due to presence of the thermal oscillations. Thus, we can make the estimation $\phi^2/L^2\sim\tfrac{\kappa_B T}{\rho_0 L^2}\sim5\cdot 10^{-3}$, hence both initial pretension and tension in the membrane due to the oscillations are important. In the following for the estimations of the effect we disregard this self-consistent dependence, setting the pretension to a given constant; all estimations will be decreased in the case of the absence of initial pretension.
\end{thebibliography}
\end{document}